\documentclass[a4paper,fleqn,usenatbib]{mnras}  

\usepackage[T1]{fontenc}
\usepackage{ae,aecompl}
\usepackage{graphicx}	
\usepackage{amsmath}	
\usepackage{amssymb}	

\newcommand{\hii}{\textrm{H}\textsc{ii}}

\newcommand{\oiidoub}{[\textrm{O}\textsc{ii}]\ensuremath{\lambda3727,3729}}
\newcommand{\oiilam}{[\textrm{O}\textsc{ii}]\ensuremath{\lambda3727}}
\newcommand{\oiiiv}{[\textrm{O}\textsc{iii}]\ensuremath{\lambda5007}}

\newcommand{\oiiidoub}{[\textrm{O}~\textsc{iii}]\ensuremath{\lambda\lambda4959,5007}}
 
\newcommand{\ha}{\ifmmode {\rm H}\alpha \else H$\alpha$\fi}
\newcommand{\hb}{\ifmmode {\rm H}\beta \else H$\beta$\fi}
\newcommand{\lya}{\ifmmode {\rm Ly}\alpha \else Ly$\alpha$\fi}
\newcommand{\pg}{\ifmmode {\rm P}\gamma \else Pa$\gamma$\fi}
\newcommand{\lyb}{\ifmmode {\rm Ly}\beta \else Ly$\beta$\fi}
\newcommand{\lyg}{\ifmmode {\rm Ly}\gamma \else Ly$\gamma$\fi}

\newcommand{\ciiidoub}{\textrm{C}\textsc{iii}]\ensuremath{\lambda\lambda1907,1909}}

\newcommand{\civ}{\textrm{C}\textsc{iv}\ensuremath{\lambda1548,1550}}

\newcommand{\heii}{\textrm{He}\textsc{ii}\ensuremath{\lambda1640}}
\newcommand{\oiiiuv}{\textrm{O}\textsc{iii}]\ensuremath{\lambda1661,1666}}

\def\kms{km s$^{-1}$}

\def\ergs{\ifmmode \mathrm{erg\hspace{1mm}s}^{-1} \else erg s$^{-1}$\fi}
\def\ergscm{erg s$^{-1}$ cm$^{-2}$}
\def\micron{\ifmmode \mu\mathrm{m} \else $\mu$m\fi}
\def\msun{\ifmmode \mathrm{M}_{\odot} \else M$_{\odot}$\fi}
\def\msunyr{\ifmmode \mathrm{M}_{\odot} \hspace{1mm}{\rm yr}^{-1} \else $\mathrm{M}_{\odot}$ yr$^{-1}$\fi}
\def\zsun{\ifmmode Z_{\odot} \else Z$_{\odot}$\fi}
\def\lsun{\ifmmode L_{\odot} \else L$_{\odot}$\fi}
\def\mstar{\ifmmode \mathrm{M}_{\star} \else M$_{\star}$\fi}

\title[Resolving parsec-scale dense star-forming regions at high-z]
{Paving the way for the JWST: witnessing globular cluster formation at $z>3$}

\author [E.~Vanzella et al.]{
\parbox[t]{\textwidth}{E.~Vanzella$^1$\thanks{E-mail: eros.vanzella@oabo.inaf.it},
F.~Calura$^1$, M.~Meneghetti$^1$, A.~Mercurio$^2$, M.~Castellano$^3$, G.~B.~Caminha$^4$, I.~Balestra$^5$, P.~Rosati$^4$,
P.~Tozzi$^6$, S.~De Barros$^7$, A.~Grazian$^3$, A.~D'Ercole$^1$, L.~Ciotti$^{8}$, K.~Caputi$^9$, C.~Grillo$^{10,11}$, E.~Merlin$^3$,
L.~Pentericci$^3$, A.~Fontana$^3$, S.~Cristiani$^{12}$, and D.~Coe$^{13}$
}
\vspace*{8pt}\\
$^1$INAF -- Osservatorio Astronomico di Bologna, via Ranzani 1, 40127 Bologna, Italy\\
$^2$INAF -- Osservatorio Astronomico di Capodimonte, Via Moiariello 16, I-80131 Napoli, Italy\\
$^3$INAF -- Osservatorio Astronomico di Roma, Via Frascati 33, I-00078 Monte Porzio Catone (RM), Italy\\
$^4$Dipartimento di Fisica e Scienze della Terra, Universit\`a degli Studi di Ferrara, via Saragat 1, I-44122 Ferrara, Italy\\
$^5$University Observatory Munich, Scheinerstrasse 1, 81679 Munich, Germany\\
$^6$INAF -- Osservatorio Astrofisico di Arcetri, Largo E. Fermi, I-50125, Firenze, Italy\\
$^7$Observatoire de Gen\`eve, Université de Gen\`eve, 51 Ch. des Maillettes, 1290, Versoix, Switzerland\\
$^8$Department of Physics and Astronomy, University of Bologna, via Ranzani 1, I-40127, Bologna, Italy\\
$^9$Kapteyn Astronomical Institute, University of Groningen, Postbus 800, 9700 AV Groningen, The Netherlands\\
$^{10}$Dipartimento di Fisica, Universit\`a  degli Studi di Milano, via Celoria 16, I-20133 Milano, Italy\\
$^{11}$Dark Cosmology Centre, Niels Bohr Institute, University of Copenhagen, Juliane Maries Vej 30, DK-2100 Copenhagen, Denmark\\
$^{12}$INAF -- Osservatorio Astronomico di Trieste, via G. B. Tiepolo 11, I-34143, Trieste, Italy\\
$^{13}$Space Telescope Science Institute, Baltimore, MD, USA\\
}

\begin{document}
\date{}
\maketitle

\begin{abstract}
  We report on five compact, extremely young ($<10$~Myr) and blue
  ($\beta_{UV}$ < $-2.5$, $F_{\lambda} =\lambda^{\beta}$) objects
  observed with VLT/MUSE at redshift 3.1169, 3.235, in addition to
  three objects at $z=6.145$.  These sources are magnified by the
  Hubble Frontier Field galaxy clusters MACS~J0416 and AS1063. 
  Their de-lensed half light radii ($R_e$) are between 16 to
  140 pc, the stellar masses are $\simeq 1-20 \times 10^{6}$ \msun, 
  the magnitudes are $m_{UV}=28.8 - 31.4$ 
  ($-17<M_{UV}<-15$) and specific star formation rates can be as large as $\sim800$ Gyr$^{-1}$.  
  Multiple images of these systems
  are widely separated in the sky (up to $50''$) and
  individually magnified by factors 3-40.  Remarkably, the inferred
  physical properties of two objects are similar to those expected in
  some globular cluster formation scenarios, representing
  the best candidate proto-globular clusters (proto-GC)
  discovered so far. Rest-frame optical high dispersion spectroscopy of 
  one of them at $z=3.1169$ yields 
  a velocity dispersion 
   $\sigma_{v} \simeq 20$~\kms, implying a dynamical mass dominated
  by the stellar mass.  Another object at $z=6.145$, with de-lensed
  $M_{UV} \simeq -15.3$ ($m_{UV} \simeq 31.4$), shows a stellar mass
  and a star-formation rate surface density consistent with the values expected from
  popular GC formation scenarios.  An additional
  star-forming region at $z=6.145$, with de-lensed $m_{UV}\simeq32$,
  a stellar mass of  0.5 $\times10^{6}$ \msun\ and a star formation rate of 0.06 \msunyr\ 
  is also identified.  These objects
  currently represent the faintest spectroscopically confirmed
  star-forming systems at $z>3$, elusive even in the deepest blank
  fields. We discuss how proto-GCs might contribute to the ionization
  budget of the universe and augment Ly$\alpha$ visibility during reionization. 
This work underlines the crucial role of JWST in characterizing the rest-frame optical
and near-infrared properties of such low-luminosity high$-z$ objects.
\end{abstract}


\section{Introduction}
Globular clusters host the most ancient stars in the Universe. 
Despite being among the most studied stellar systems and 
despite the existence of a large variety of models for their formation, 
the initial conditions and the physical processes driving their growth 
and characterising their earliest evolutionary phases are yet to be understood. 
Nowadays, it has become clear that GC can no longer be regarded as a simple stellar population 
(SSP), i.e. an assembly of coeval stars sharing identical chemical composition. 
Over the last decade, substantial evidence has been gathered in favour of 
the presence of multiple stellar populations within globular clusters 
\citep{gratton04, piotto07,dantona08}. 

On the theoretical side, much effort has been devoted to the 
development of new models for the formation and evolution of globular clusters.
In order to take into account the existence of multiple stellar populations, 
most popular models for GC formation 
consider a scenario in which a second generation (SG) of stars forms from the gas
ejected by either first generation (FG) asymptotic giant branch (AGB) stars
\citep{cottrell81,dantona04,dercole08}, 
or FG fast rotating massive stars \citep{prantzos06, decressin07}, 
as well as massive interacting binaries \citep{demink09}, or 
even supermassive ($M>10^4 M_{\odot}$) stars \citep{denissenkov14}. 
 
The predominance of SG stars observed today in most GCs implies a
problem which is common to all the scenarios invoking a standard
stellar initial  mass function (IMF) for FG stars, i.e. the
so-called 'mass-budget' problem.  In fact, for a \citet{salpeter55} or
a \citet{kroupa01} IMF, the gas shed by the massive stars during the
H-burning phase, or the gas contained in the envelopes of massive AGBs
\citep{dantona04,dercole08,renzini15}, is in general too scarce to
form a large SG population. One can solve this problem by postulating
that the GC precursors are {\it more} massive (by factors between 5
and 20) than the objects we observe today \cite[e.g.,][]{dercole08,
  renzini15}, or alternatively, by invokig a highly non-standard IMF
for the FG, particularly rich in massive stars \citep{dantona04,
  downing07}. The latter scenario, however, faces some serious difficulties
when considering the maximum amount of helium which can be produced by
a FG of super massive stars \citep{renzini15}.  
A scenario in which
proto-GCs are more massive than today GCs opens the interesting
possibility to detect and spatially resolve their stellar emission.
 
 \begin{figure*}
\centering
\includegraphics[width=17cm]{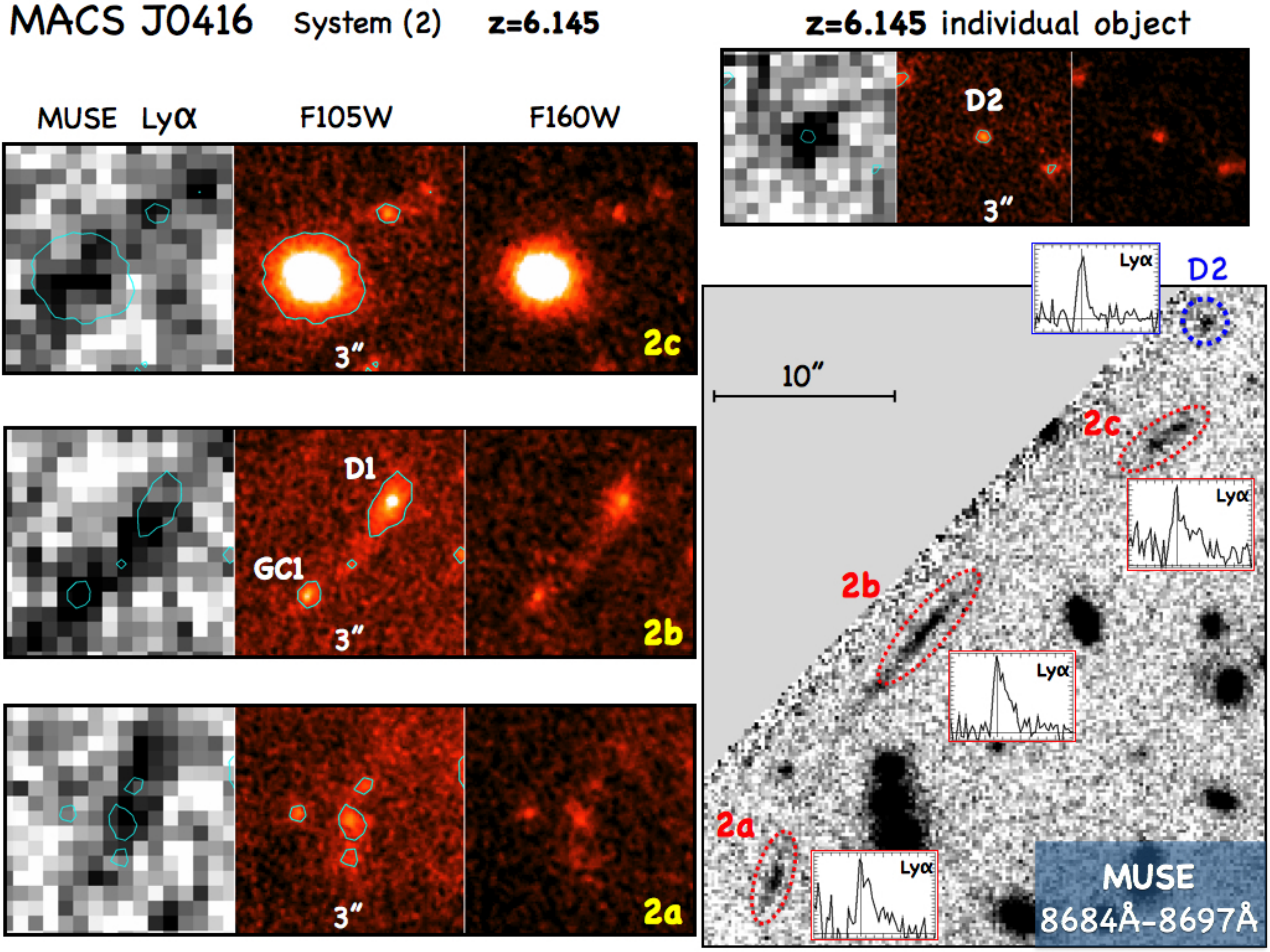}
\caption{The giant Lyman-alpha arc in MACS~J0416 (45$''$ wide) formed by the multiple images 2a, 2b and 2c 
and the second object (D2) at $z=6.145$ are shown in the HST near infrared bands (stellar continuum)
and MUSE data-cube (Ly$\alpha$). The insets show the one-dimensional Ly$\alpha$ profiles.}
\label{arc6}
\end{figure*}
 
The main scenarios proposed so far to explain multi-population GCs
postulate the formation of FG stars at $z>2$
\cite[e.g.,][]{kruijssen15, dercole16}.  The formation of GCs has also been studied in the context of cosmological models, which predict their 
birth within high-density regions of galactic discs at $z>2$
\cite[e.g.,][]{kravtsov05}, or envisage that FG stars are formed in an
intense burst in the center of a minihalo during major merger events
\cite[][]{trenti15}, or consider their formation at the center of
primordial dwarf galaxies \cite[][]{ricotti16}.

The direct observation of the formation of FG stars within
high-redshift proto-GCs would provide invaluable clues on how to
disentangle such a range of GC formation scenarios.

On the observational side, over the last years many studies have been
carried out to derive the faint-end slope of the ultraviolet galaxy
luminosity function, its faint-end cut-off $M_{lim}$, and the Lyman-continuum photon production
efficiency \cite[e.g.,][]{bouwens16a}.  A major uncertainty in the interpretation of these
observations is represented by the observed fraction of ionizing
radiation that escapes from galaxies and reach the intergalactic and
circum-galactic media \cite[see e.g.,][]{grazian16, siana15, vanz12b}.
Irrespective of the nature of the contributors, the general consensus is that the faintest objects
represent the main producers of the ionizing  background at high redshift \citep{wise14,bouwens15,ferrara13,kimm14}. 
Noteworthy, proto-GCs may also represent good candidates for 
the sources of radiation which reionized the universe by $z\sim6$ \cite[e.g.,][]{ricotti02, schaerer11, ricotti16}.

The detection of extremely faint galaxies at high redshift has been
reported in the literature, with magnitudes as faint as 
$M_{UV} \simeq -14$ both at redshift $2-3$ \cite[e.g.,][]{alavi14,alavi16} and $6-8$
\cite[e.g.,][]{atek15, livermore16, bouwens15, vanz14, bouwens16c}.  These systems
are characterised by very low stellar masses, of the order of a few
$10^{6}$ solar masses \cite[e.g.,][]{karman16}.  A still open question
concerns the nature of the objects belonging to this luminosity
domain: are we dealing with dwarf galaxies \cite[e.g.,][]{finlator17}, \hii\ galaxies \cite[e.g.,][]{terlevich16}, 
super star-clusters, or extremely compact star
clusters or clumps? (i.e., with sizes of the order of a few tens of
parsec, \citealt{bouwens16b,kawamata15, livermore15, vanz16a, ellis01}).  
In order to
answer this question, we need to derive a few basic physical
quantities such as the stellar mass, the star-formation rate, as well
as the size of these systems. 

Currently, deep and gravitationally lensed fields observed with the
Hubble Space Telescope offer an opportunity to measure such physical
properties, even for extremely faint sources.
Much progress has recently being driven by deep observations of
massive galaxy clusters, carried out in the context of large HST
programs, particulalry the Hubble Frontier Fields (HFF) survey
\citep{lotz14,lotz16, koekemoer14}.  Exploting gravitational lensing, cluster
cores are used as cosmic telescopes to look deeply into the distant
universe. High-precision lens models can be built using a large number
of multiply lensed sources spanning a large redshift range, which
however need to be spectrosopically identified with ground-based or
HST grism observations \citep{treu15}. In particular, in
combination with the very efficient integral field spectrograph MUSE
on the VLT \citep{bacon10} \footnote{{\it www.eso.org/sci/facilities/develop/instruments/muse.html}}, 
the identification and characterization of Lyman-$\alpha$
emitting galaxies near the flux limit of the Hubble imaging data, in
lensed and blank fields, has become possible \cite[e.g.,][]{karman16,
  cam16c, vanz16a}.

In this work we study 
five systems detected behind two Hubble Frontier fields at redshift
3.1169, 3.235 and three at 6.145 extracted from deep MUSE observations
of \citet{karman16}, \citet{vanz16a} and \citet{cam16c}.
Specifically, we provide novel estimates for the size, dynamical mass
and SED-fitting using the full-depth HFF photometry and near-infrared
spectroscopy for the object at $z=3.1169$ behind the galaxy cluster
AS1063 (named ID11, also discussed in \citet{vanz16a,karman16}).  The
redshift of the remaining four objects have been presented in
\citet{cam16c} and used to constrain the lens model of
MACS~J0416. Here we focus on their physical properties.  We also
explore the possibility that some of these sources may represent
proto-GC caught during the formation of their stellar FG.  The lens
models of the two galaxy clusters discussed in this work have been
constrained using dozens of multiple systems spectroscopically
confirmed at $3<z<6.5$ with MUSE \citep{cam16a, cam16c}.

The paper is structured as follows: in Sect. 2 the strongly magnified
systems are presented together with the lens models and the inferred
sizes. Sect. 3 describes the photometry and the SED-fitting used to
derive the physical quantities. We discuss the results in Sect. 4 and
conclude in Sect. 5.

In this work, we assume a flat cosmology with $\Omega_{M}$= 0.3,
$\Omega_{\Lambda}$= 0.7 and $H_{0} = 70$ km s$^{-1}$ Mpc$^{-1}$,
corresponding to 7650, 7560 and 5560 physical parsec for $1''$
separation at redshift 3.1169, 3.235 and 6.145, respectively.

\section{Extremely small sources in the Hubble Frontier Fields}

The spatial investigation of extremely compact stellar systems ($<100$pc) at $z > 3$ is precluded 
in field surveys (as also demonstrated by dedicated simulations presented in 
Appendix~\ref{simula}, see also \citealt{bouwens16b}). As an example, an object with a  proper 
half light radius of 50 pc at $z=3(6)$ corresponds to 7(9) mas in the sky, or 0.30 (0.23) pixels, assuming 1 pixel=$0.03''$ 
(typically the  spatial scale of drizzled HST images).  Strong gravitational lensing
allows us to probe physical sizes as small as 20-60 parsec  at $z>3$. This is presented in the next sections.


\subsection{The strongly lensed systems}
Over one hundred multiple images have been identified behind the
Hubble Frontier Fields galaxy clusters AS1063 and MACS~J0416. For the
majority of them, the confirmation of their lensing origin has been
possible through MUSE observations \citep{karman16,cam16a,cam16c}. In
particular MUSE integral field spectroscopy has revealed secure
Ly$\alpha$ emission from widely separated multiple images. We focus on
three systems of multiple images confirmed at redshift 3.1169, 3.235
and 6.145.  The angular separation of the multiple images in these
systems extends to 50$''$, $20''$ and $45''$, respectively.  These systems are selected
on the basis of their strong magnification and high signal-to-noise detection in the HST
images (S/N $>10-30$). This enables a solid characterization of their sizes and a
measurement of their physical properties from SED fitting.

\begin{figure*}
\centering
\includegraphics[width=16cm]{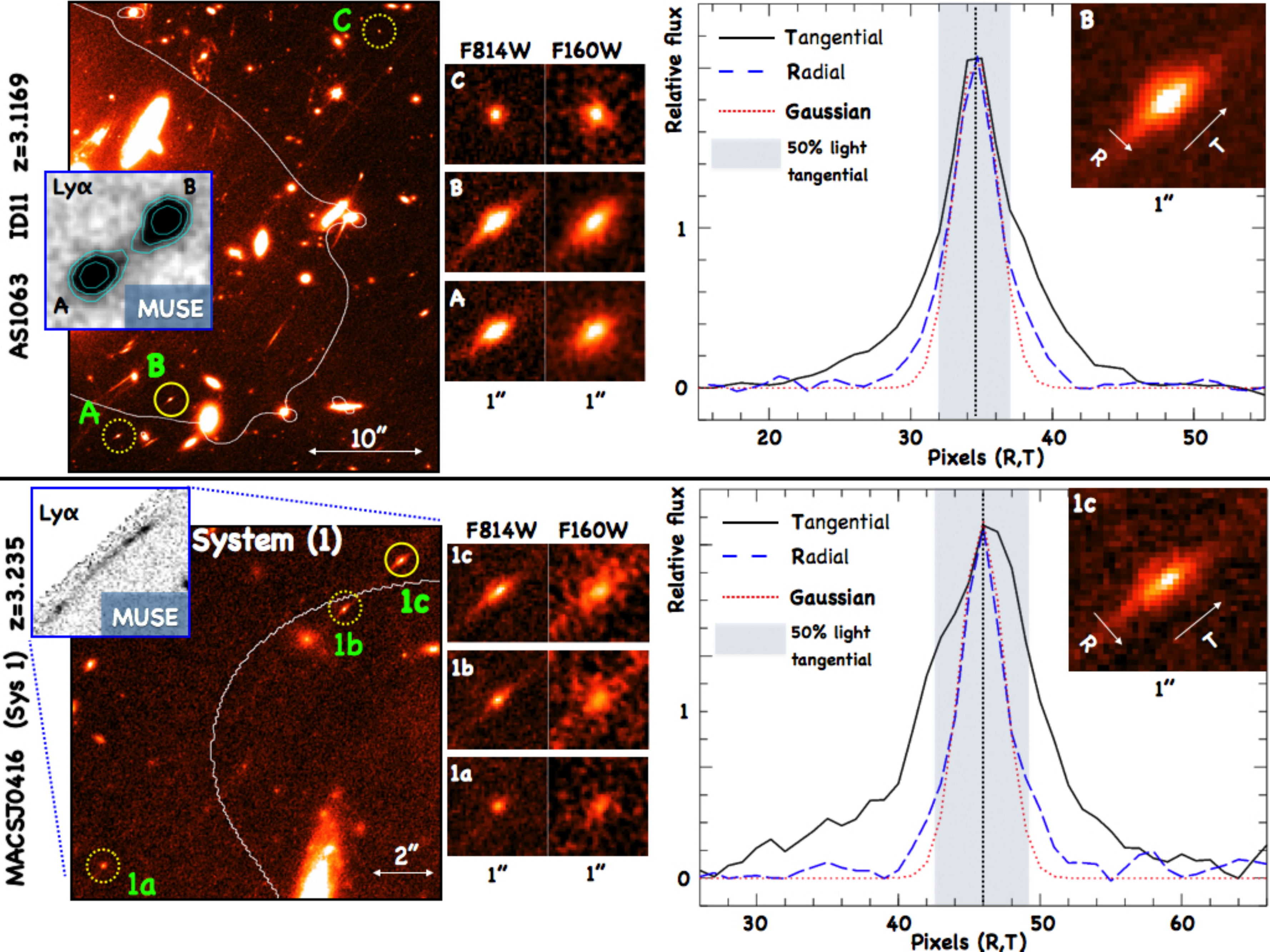}
\caption{Multiple images and observed light profiles along tangential
  and radial directions for two sources at redshift 3.1169 and 3.235,
  in AS1063 and MACJ0416, respectively.  On the left side the multiple
  images of each system are shown in the F814W band, including the
  two-dimensional Ly$\alpha$ emission from VLT/MUSE (insets). The thin
  white lines represent the critical curves at the redshift of the
  objects.  The zoomed multiple images are also shown in the F814W and
  F160W bands. 
  On the right side, the spatial profiles and images are shown for the most magnified images 
  (image 1c, bottom-right, and image B, top-right).  
  The object is well resolved along the tangential direction (T), whereas it is not (or barely)
   resolved along the radial direction (R).}
\label{arcs3}
\end{figure*}

Several images appear distorted tangentially  by the cluster mass
distribution (as shown in Figures~\ref{arc6} and ~\ref{arcs3}),
indicating that they are close to the cluster tangential critical
lines \citep[see e.g.][]{kneib11}. In this case, the magnification is
dominated by the tangential component ($\mu_{T}$), defined as
\begin{equation}
\mu_T=(1-\kappa-|\gamma|)^{-1} \;,
\end{equation}
where $\kappa$ is the dimensionless surface-density (or convergence) and $\gamma$ is the shear.  
Thanks to this magnification factor, the sources investigated here are spatially resolved along the tangential direction. 
On the contrary, the radial component of the magnification,
\begin{equation}
\mu_R=(1-\kappa+|\gamma|)^{-1} \;,
\end{equation}
is sub-dominant at the position of the images.  The light profile along the radial direction is consistent with the PSF, indicating
that the sources are radially un-resolved (or marginally resolved at most). The total
magnification is $\mu_{tot} = \mu_{R} \times \mu_{T}$. The estimated
magnifications for the cases studied in this work are reported in Table~\ref{infos} and discussed in detail below. 

To estimate the half light radius of the sources ($R_e$), we measure the light profiles along the tangential direction. Accounting for
the tangential magnification, we derive a de-lensed value of $R_e$. In this process, we assume that the sources are intrinsically
circular, and that the observed tangential elongation is only due to magnification. 

 \begin{figure*}
\centering
\includegraphics[width=15cm]{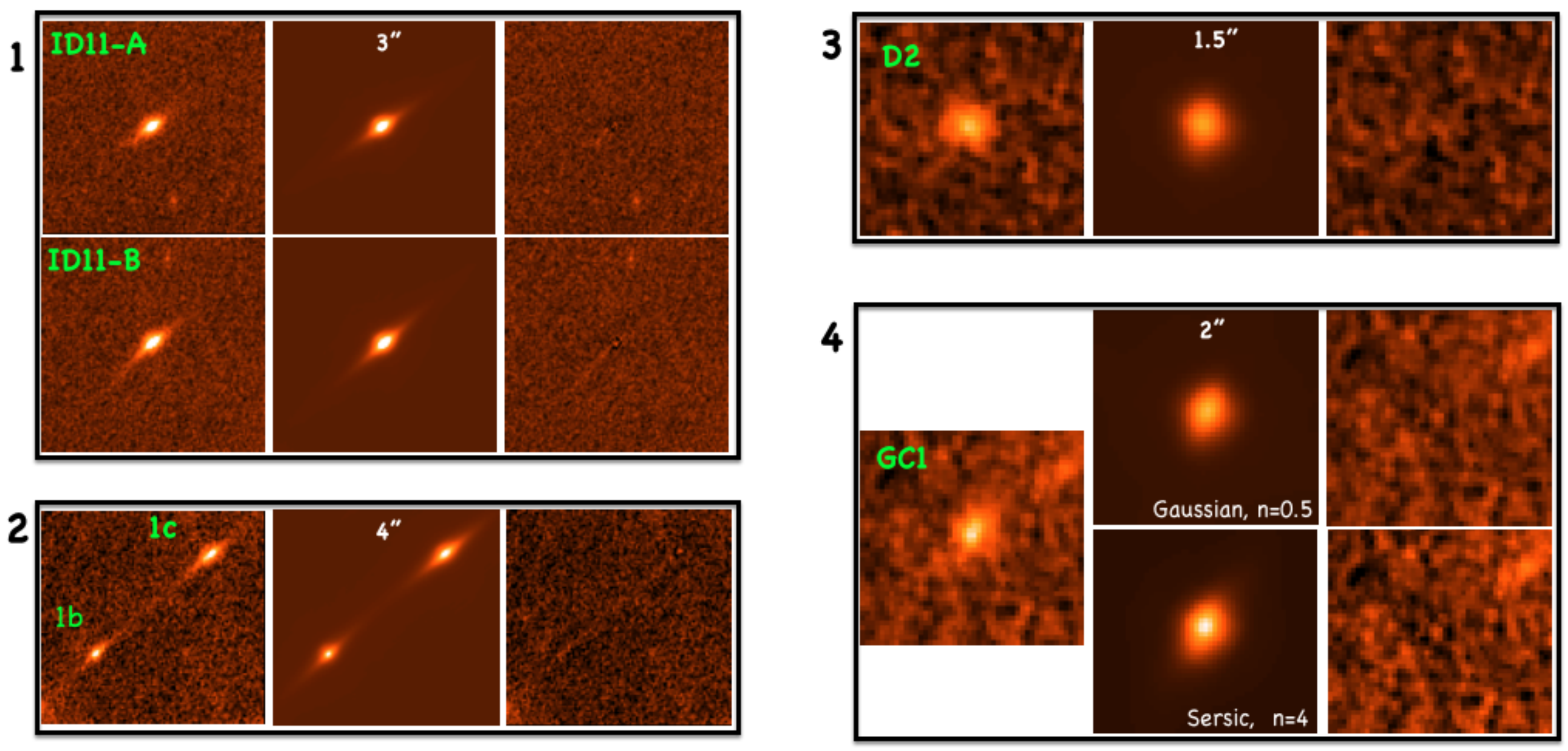}
\caption{{\tt Galfit} fitting of the four compact and most magnified objects in this study:
ID11-A and B at $z=3.1169$ (panel 1), the system 1c,b at $z=3.235$ (panel 2), and D2 (panel 3) and GC1 (panel 4) at $z=6.145$. 
The observed images, {\tt Galfit} models and residuals
(observed-models) are shown from left to right for each system. The morphological parameters are reported in Table~\ref{morph}.}
\label{galfit}
\end{figure*}

The circularized effective radius is also calculated (when possible)
as $R_{c} =R_{e} q^{0.5}$ where $q=a/b$ is the axis ratio between the
minor and major axes of the source. Such parameter gives an upper
limit when the radial component is not resolved.

In the rest of the work, the effective radii, expressed in parsec, and
the physical quantities are always intrinsic (i.e., de-lensed),
whereas radii reported in pixels are by definition observed quantities
(1 pixel corresponds to 0.03$''$).


\subsection{System ID11 in AS1063 at $z=3.1169$}
 
We revisit here the physical size of the object at $z=3.1169$ studied
by \citet{vanz16a} and perform SED-fitting using the full depth HFF
photometry.  Among the sources presented in this work, this is the
system with the highest signal-to-noise ratio in the HFF photometry
(ranging between 20 and $>50$ for the three multiple images).
While the separation between images A and C is very large 
($\simeq 50''$), images A and B are closer, $\lesssim 10''$, and presumably
positioned on opposite sides of the cluster tangential critical line
(see the white curve in Figure~\ref{arcs3}). Based on the recent
analysis by \citet{meneghetti16}, it is expected that the uncertainty
on the magnification estimated from lens models is a steep function of
the magnification itself, being $\gtrsim 50\%$ for $\mu>10$. Although
this condition applies to images A and B, we can obtain a more robust
estimate of the magnification of these two images based on the
following considerations.  In such a ``fold'' image configuration, it
is expected that the two images have similar magnifications and
inverse parity.  Indeed, images A and B have very similar shapes and
fluxes ($f_{B}/f_{A} \simeq 1.1$).  Among the three images, the
faintest one, C, has the least uncertain magnification factor and flux ratio
$f_{B}/f_{C} = 4.0\pm0.05$. These ratios have been inferred by including all
the HST/ACS bands and measuring the average flux ratios among them.
In this work, we revise the model of  \cite{cam16a} in order to optimally reproduce  
the observed positions {\sl and} flux ratios of the three images of ID11. With this model, we infer a
a magnification  $\mu_{C} = 5.0
\pm 0.2$.  The magnification of the counter-images A and B is
derived from the observed flux ratios between C and images A, B as
in \citet{vanz16a}.  The resulting total magnifications for images A
and B are $\mu_{tot}^{A}=18.2$ and $\mu_{tot}^{B}=20.0$
respectively, 
with errors smaller than 10\%.  

As already stated, the images A and B are dominated by
the tangential magnification ($\mu_{T}$). Indeed, the radial
magnification estimated by the model is quite similar for all three
images ($\mu_{R}\simeq 1.3$), while the tangential magnifications are
$\mu_{T}\simeq 16$ and 15 for images B and A, respectively.  As shown in Fig.~\ref{arcs3}, the light profile along the radial direction is consistent with the PSF, and it is thus unresolved. On the contrary, we can spatially
resolve the core and the light profile of the source in the tangential direction. 

In order to measure the intrinsic (i.e. un-lensed)  size of ID11, we use the {\tt Galfit} software \citep{peng02, peng10} to fit the sources in  both images
in the F814W band (probing 2000\AA~rest-frame) and F160W band (probing 4000\AA~rest-frame). Then, we use the model  magnification to obtain the de-lensed sizes.
The relevant parameters are reported in Table~\ref{morph}. 
The tangentially lensed  effective radii measured in the F814W (2000\AA) and F160W (4000\AA) bands
turn out to be very similar, $R_e$(2000\AA) $ = 3.0 \pm 0.5$ and $R_e$(4000\AA) $ = 4.0 \pm 1.0$ pixels, respectively.
This corresponds to $R_e=43 \pm 7$ pc and $57 \pm 14$ pc, respectively ($1''=7650$pc at $z=3.1169$)  after de-lensing.
The de-lensed circularized effective radius is $R_c = R_e \times q^{0.5} \simeq 50$ pc. 

Fig.~\ref{arcs3} shows that the light profile is far from a simple Gaussian shape, rather it  turns out to follow a S\'ersic profile with
index $n \simeq 4$ (Figure~\ref{galfit}).  In addition, the light profiles  are quite symmetric, despite the gradient of the
magnification along the tangential direction is large (being close to
the critical line).  This further supports the intrinsic compactness
of the source. In fact, a lensed, more extended object in the source
plane would  deviate from the symmetric profile generating a
boosted tail in the direction of the critical line, towards which the
tangential magnification increases rapidly. In our case, the observed
stellar continuum follows a symmetric profile instead, showing that
differential magnification across the image is
negligible. 
On the contrary, this effect is detected in the two-dimensional Ly$\alpha$
emissions of images A and B. Figure~\ref{arcs3} shows the Ly$\alpha$
emissions averaged  over 5 slices in the MUSE data-cube (see also
\citealt{karman15}).  In this case the asymmetric shape is observed as
two ``asymmetric-lobes'' pointing towards the critical line in the
middle (see the inset of Figure~\ref{arcs3}).  This suggests the
Ly$\alpha$ emission arises from a region that is larger than the
stellar continuum measured in the F814W band, which almost touches the
lens caustic  on the source plane.

In Appendix~\ref{simula}, we describe a set of end-to-end simulations which validate our  
method to derive intrinsic radii based mainly on the tangential
magnification.  Simulations show that the structural properties of the sources are recovered using
our procedure and provide further evidence supporting our conclusions on the size and on the brightness 
profile of ID11. 

\begin{table*}
\footnotesize
\caption{Intrinsic physical properties and magnifications predicted for the most magnified image of each system (see text for details). 
Column 1 shows the typical physical quantities expected from the AGB  scenario of GC formation \citep{dercole08}. Columns $3-7$ report
the inferred properties for the five objects considered in this work: GC1, D1, D2 and Sys\_1 identified by \citet{cam16c}, and ID11 discussed in
\citet{vanz16a}.  In square brackets the 68\% c.l. is reported (additional constraints from the \lya\ line can decrease significantly these
uncertainties, see Sect.~\ref{sedfitting}).  De-lensed apparent and absolute magnitudes are reported. 
$f(+)/f(-)$ is the flux ratio between the most and the least
magnified images within the same system. (*) Astrodeep IDs \citep{castellano16b,merlin16}.  (**) These are half mass radii; they must be 
reduced by a factor 1.33 to obtain the two-dimensional half light radii \citep{wolf10}.  The age is the elapsed time since the onset of SF.}
\begin{tabular}{l l c c c c c} 
\hline
                                                            &                                & GC1(z=6.145)                  &D1(z=6.145)                              &D2(z=6.145)                        & Sys\_1c(z=3.235)                    & ID11\_B(z=3.1169)\\ 
                                                            &                                &   (*)2169                            &   2179                                         &      2411                               & 2268                                            & -- \\ 
                                                            &   Models                &  04:16:11.56                     &  04:16:11.48                             &04:16:10.31                         & 04:16:11.15                               & 22:48:41.56 \\ 
                                                            &   Proto-GCs          & -24:03:44.7                       &-24:03:43.4                                 &-24:03:25.8                         & -24:03:37.4                               & -44: 32:23.9\\                                                             
\hline
Stellar mass [$10^{6} M_{\odot}$] &  1 , 10                   &  $68_{[21,3273]} \mu_{tot}^{-1}$ &$380_{[368,585]}\mu_{tot}^{-1}$ & $16_{[12,1027]}\mu_{tot}^{-1}$& $55_{[43,848]}\mu_{tot}^{-1}$   & $400_{[280,560]}\mu_{tot}^{-1}$ \\
SFR [$M_{\odot} yr^{-1}$]               &  0.2 , 2.0               &    $54_{[1,165]}\mu_{tot}^{-1}$& $275_{[131,585]}\mu_{tot}^{-1}$& $5_{[0.5,48]}\mu_{tot}^{-1}$     & $10_{[2,12]}\mu_{tot}^{-1}$          & $14_{[7,20]}\mu_{tot}^{-1}$ \\ 
Age [Myr]                                           &  5                            &   $1.3_{[1,708]}$                 & $1.4_{[1,3]}$                           & $3.2_{[1,710]}$                    &  $5.6_{[4.5,500]}$                    & $13_{[2,60]}$ \\
E(B-V)                                                &  $\simeq 0$          &    $\lesssim 0.15$               & 0.10                                           & 0.0                                          &  0.06                                         & 0.0 \\
$R_e$ (UV)  [pc]                               &  16 , 35(**)            &   $16\pm7$                         &$140\pm 13$                            & $<100$                                 & $80-45$                                   & $43\pm7$\\   
$R_c$ (UV)  [pc]                                     &  $''$                        &   $<30$                                &$150\pm 20$                             &$<100$                                  & $<(80-45)$                               & $\lesssim50$\\   
$\Sigma_{SMD}$ [$M_{\odot} pc^{-2}$] & 800$-$1720 & $1400_{-900}^{+2400}$  & $295_{-80}^{+100}$                & $>85$                                    &   $36-39$                                     & $1300_{-500}^{+750}$ \\  
$\Sigma_{SFR}$ [$M_{\odot} yr^{-1} pc^{-2}$] & (1.6-3.4)$10^{-4}$ & $2.7\times 10^{-3}$ & $2.3\times 10^{-4}$ & $>5.3\times 10^{-5}$ & $(1.2-1.4) 10^{-5}$ & $1.2\times 10^{-4}$\\
\hline
m(1500\AA)                                      &  $\simeq 29 - 32 $ &   $31.4 \pm 0.2$     &  $29.7  \pm 0.2$    &  $29.6\pm 0.3$ & $30.9-32.1$          & $28.8\pm 0.1$ \\
M(1500\AA)                                      &  $ > -17 $                &   $-15.3$                  & $-17.0$                   &  $-17.1$              & $-15.3, -14.1$   & $-17.1$ \\
$\beta_{UV}$                                   &   $\lesssim -2.5$     & $-2.52\pm0.36$     &   $-2.40\pm0.16$  & $-2.85\pm0.43$ &  $-2.64 \pm 0.15$ & $-2.75\pm0.20$\\
\hline
$\mu_{tot}$                                       &  --                             &    $25.0 \pm 2.5$      & $19.0 \pm 2.0$  &  $3.0 \pm 0.2$    &  $37-110$                  & $ 20.0 \pm 2.1$ \\ 
$\mu_{tang}$                                   &  --                             &    $17.5 \pm  2.0$     & $13.4 \pm 1.5$   &  $1.7 \pm 0.1$  &  $19-33$                    & $ 16.2 \pm 1.6$\\ 
$f(+)/f(-)$                                           &  --                             &    $\simeq 2.5$         & $\simeq 2.5$      &  --                         &   $3.7\pm0.1$        & $4.00\pm0.05$\\
\hline
\hline 
\end{tabular}
\label{infos}
\end{table*}

\subsection{System 1 in MACS~J0416 at $z=3.235$}

This strongly lensed object is identified as System (1) (Sys\_1) in \citet{cam16c}. 
The bottom-left panel of Figure~\ref{arcs3} shows the Ly$\alpha$ emission as a continuous arc-like shape at $z=3.235$, captured
in the MUSE data-cube, and the three multiple images 1a, 1b and 1c generating such a line emission
(extending up to $\simeq 17''$). 
Also in this case, all the multiple images are well detected in the HST images with $S/N\simeq10-50$ and the geometry of the system
is very similar to ID11. We focus on
the tangential direction which  allows us to resolve and extract firm constraints on the physical size of the core of the
object. In particular, following the same methodology discussed above, we start from the least  magnified
image 1a  and rescale it to the observed flux ratio $f_{1c}/f_{1a}=3.7 \pm 0.1$ ($\simeq \mu_{1c} / \mu_{1a}$). 
This ratio was inferred by including all the HST/ACS bands and measuring the average flux ratio among them.
While the direct estimates of $\mu_{1b}$ and $\mu_{1c}$
are affected by large errors, image 1a is the less magnified and the more stable among the others.
From the lens model of \citet{cam16c} we derive $\mu_{1a}=30 \pm 15$, and from 7 different models available from the
HFF lens tool calculator\footnote{{\it https://archive.stsci.edu/prepds/frontier/lensmodels/}} 
its median value is $<\mu_{1a}> \simeq 15 $ (all models providing a magnification higher than 10).
Therefore, the total $\mu_{1c}$, after rescaling by the flux ratio ($\simeq 3.7$), ranges between 37 and 110 for values of $\mu_{1a}$
between 10 and 30. 
However, even adopting the lowest $\mu_{1a}=10$, the resulting total magnification for image 1c is $\mu_{1c}>37$. 
Also in this case, the quite elongated shape visible in the 1c image and the overall Ly$\alpha$ arc suggest
that the magnification is mainly tangential. As discussed in the case of ID11, this is evident from the
measured $q=b/a=0.10 \pm 0.015$ of the image 1c.
The predicted tangential magnification for image 1c is therefore large and spans the range $\mu_{T}=19-33$,
depending on the total magnification assumed (37$-$110) and assuming that the source is circular. 
Figure~\ref{arcs3} and Table~\ref{infos} show the result of the {\tt Galfit} fitting, which provides a de-lensed tangential
effective radius of $R_e=6.6 \pm 2.1$ pixels. This corresponds to a physical size of 45 (80) pc adopting the tangential magnification
$\mu_{T}$=33(19). The object is detected also in the F160W-band and shows an elongated shape, however the inferred $R_e$ is strongly affected
by the low S/N ratio and a tentative estimate ranges between 70$-$150 pc (see inset of Figure~\ref{arcs3}, rest-frame 4000\AA).

\subsection{System 2 in MACS~J0416 at $z=6.145$}

\subsubsection{The giant arc}
Figure~\ref{arc6} shows the Ly$\alpha$ emission arising from a giant arc extending for $\sim 45''$ on sky and
composed by three multiple images  2a, 2b and 2c (system 2 of \citealt{cam16c}).
The large extension of the arc and the geometry of the multiple images suggest that the source is near the cusp of the lens tangential caustic and that the tangential magnification is again predominant. As expected in a cusp system, image 2b is the most magnified among the three images. Its median magnification among  the eleven lens models included in  the HFF lens tool calculator is 
$\mu_{tot} \simeq 30$. The estimates  from all models span the interval $10\le \mu_{tot} \le  80$.
However, these models have been generated without including this system (and many others now available from
\citealt{cam16c}) as constraints.  As shown in Fig. 24 of \citealt{meneghetti16}, uncertainties on image  magnifications 
are strongly reduced if nearby multiple images are used as constraints for the lens model. 
Therefore, we use the HFF models only as an estimate of the systematic uncertainties and we adopt the specific modeling 
presented in \citet{cam16c}, in which system 2 has been included and the positions of its multiple images are reproduced. 
Table~\ref{infos} reports the total magnifications, together with the tangential and radial components for image 2b.
While $\mu_{R}$ is modest and quite stable among the images ($\simeq 1.2-1.4$) and does not allow us to spatially
resolve the emitting sources in the radial direction,  the diffuse stellar emission of 2b 
appears very elongated tangentially,  as the Ly$\alpha$ emission detected in the MUSE data-cube (Figure~\ref{arc6}). 

In the following, we focus on the most magnified of the images, 2b.
We identified two distinct (sub)systems in this image, named D1 and
GC1, separated by $1.7''$. This angular separation corresponds to a
physical de-lensed separation of 0.6 kpc. GC1 is the most compact
among the two, though slightly spatially resolved along tangential
direction (see Figure~\ref{arc6}). The two objects are also present in
the Astrodeep photometric catalog with IDs 2179 and 2169, respectively
\citep{castellano16b}. Their physical properties are reported in
Table~\ref{infos} and discussed in Sect.~3.  The identification of
these two sources in the other multiple images of the arc, 2a and 2c,
is difficult, since they are very faint (expected to be $\gtrsim
1.0$ magnitudes fainter than 2b) and possibly contaminated by lower
redshift cluster members.  For example the expected magnitudes of
images 2a and 2c of GC1 are $\gtrsim 29.2$, while D1 is brighter and
possibly identified, though contaminated by a galaxy cluster member
(see Figure~\ref{arc6}).

Given the complex identification of the sub-systems GC1 and D1 in the
HST images, we use the Ly$\alpha$ fluxes detected in the MUSE
data-cube as a proxy for the relative ratios among magnifications. We
assume that the observed multiple Ly$\alpha$ lines arising from 2a,
2b, and 2c (see Figure~\ref{arc6}) probe the same lensed structure
(e.g., the sum of GC1 and D1).  The Ly$\alpha$ flux has been measured
on three apertures defined on top of the Ly$\alpha$ emissions (2a, 2b
and 2c) following the curvature of the arc and calculated collapsing
six spectral elements of 1.25\AA~each, as the best S/N estimate (and
corresponding to $\Delta v \simeq 300$ \kms\ at $z=6.145$).  The
resulting ratios are
Ly$\alpha$(2b)/Ly$\alpha$(2a)$\simeq$Ly$\alpha$(2b)/Ly$\alpha$(2c)$\simeq
2.5 \pm0.7$, and provide an estimate of the relative magnifications
between 2b and 2a,c.  The lens model of \citet{cam16c} reproduces the
positions and magnitudes of the three multiple images.  In particular
the total and tangential magnifications for objects GC1 and D1
calculated for image 2b (i.e. in the most magnified image) are
reported in Table~\ref{infos}.  Interestingly, the de-lensed
magnitude of GC1 is 31.40 in the F105W band ($27.88 \pm 0.08$ observed, \citealt{castellano16b}), 
the faintest spectroscopically confirmed object currently known at this redshift,
and fainter than any source detected in the current deepest fields
(e.g., the Hubble Ultra Deep field, \citealt{beckwith06}).

%
\begin{table*}
\footnotesize
\caption{Observed morphological parameters derived with {\tt Galfit} for the most magnified images. Morphological parameters
have been derived in the F814W and F105W bands for redshift 3 and 6, respectively, probing 2000\AA\ and 1500\AA. (*) Gaussian fit, 
object marginally resolved.}
\begin{tabular}{l c c c c c} 
\hline
Morph. parameters                           & GC1(z=6.145)         &D1(z=6.145)            &D2(z=6.145)       & Sys\_1c(z=3.235)     & ID11\_B(z=3.1169)\\ 
                                                             &   2169                        &   2179                      &      2411               & 2268                          & --  \\ 
                                                             &  04:16:11.56             &04:16:11.48            &04:16:10.31         & 04:16:11.15              & 22:48:41.56 \\ 
                                                             & -24:03:44.7               &-24:03:43.4             &-24:03:25.8          & -24:03:37.4               & -44: 32:23.9\\                                                             
\hline
$R_e$ [pix]                                                &  $1.7\pm0.7$             &$8.3\pm0.5$           &$\lesssim1.5$(*)  &  $6.6 \pm2.1$          & $3.0\pm0.5$ \\  
$q=b/a$                                               &  $<0.15$                     &$0.20\pm0.02$      &$0.3\pm0.2$        &  $<0.15$                   & $0.10 \pm 0.02$\\  
$R_c$ [pix]                                                &  $<0.70$                     & $3.71\pm0.23$     & $\lesssim 1.0$    &  $<2.5$                     & $0.95 \pm 0.17$\\  
 PA [deg]                                             &  $-39.0\pm3.0$         &$-28.5\pm0.6$        &$81\pm10$          &  $-47.0 \pm 2.0$      & $-49.59 \pm 0.5$\\   
$n$(S\'ersic)                                         &  0.5-8                          &$3.0\pm0.3$            &0.5                         & $8.0 \pm 2.0$          & 4.0-5.0\\   
mag                                                      &  $>27.3$                    &$26.1\pm0.2$       &$27.8\pm0.8$    & $26.9 \pm 0.1$    & $25.3\pm0.1$ \\   
\hline
\hline 
\end{tabular}
\label{morph}
\end{table*}

\subsubsection{A faint and dense stellar system at $z=6$: GC1}
The large tangential magnification allow us to resolve the one-dimensional
half-light radius. To this aim, we measure the FWHM of the light profiles in the F105W band (probing the
rest-frame 1500\AA) along the tangential direction, both on the image directly and by using {\tt Galfit} modeling.
Here we focus on GC1, which is the smallest of our objects.
It is the faintest object currently confirmed at this redshift with a de-lensed magnitude of 31.4 (at 1500\AA~rest-frame) 
and with a reasonably high S/N ratio in the photometry, also showing  a rest-frame Ly$\alpha$ equivalent width of the
order of 100\AA\  (correcting for the UV slope, see Table~\ref{infos}).

\begin{figure*}
\centering
\includegraphics[width=16cm]{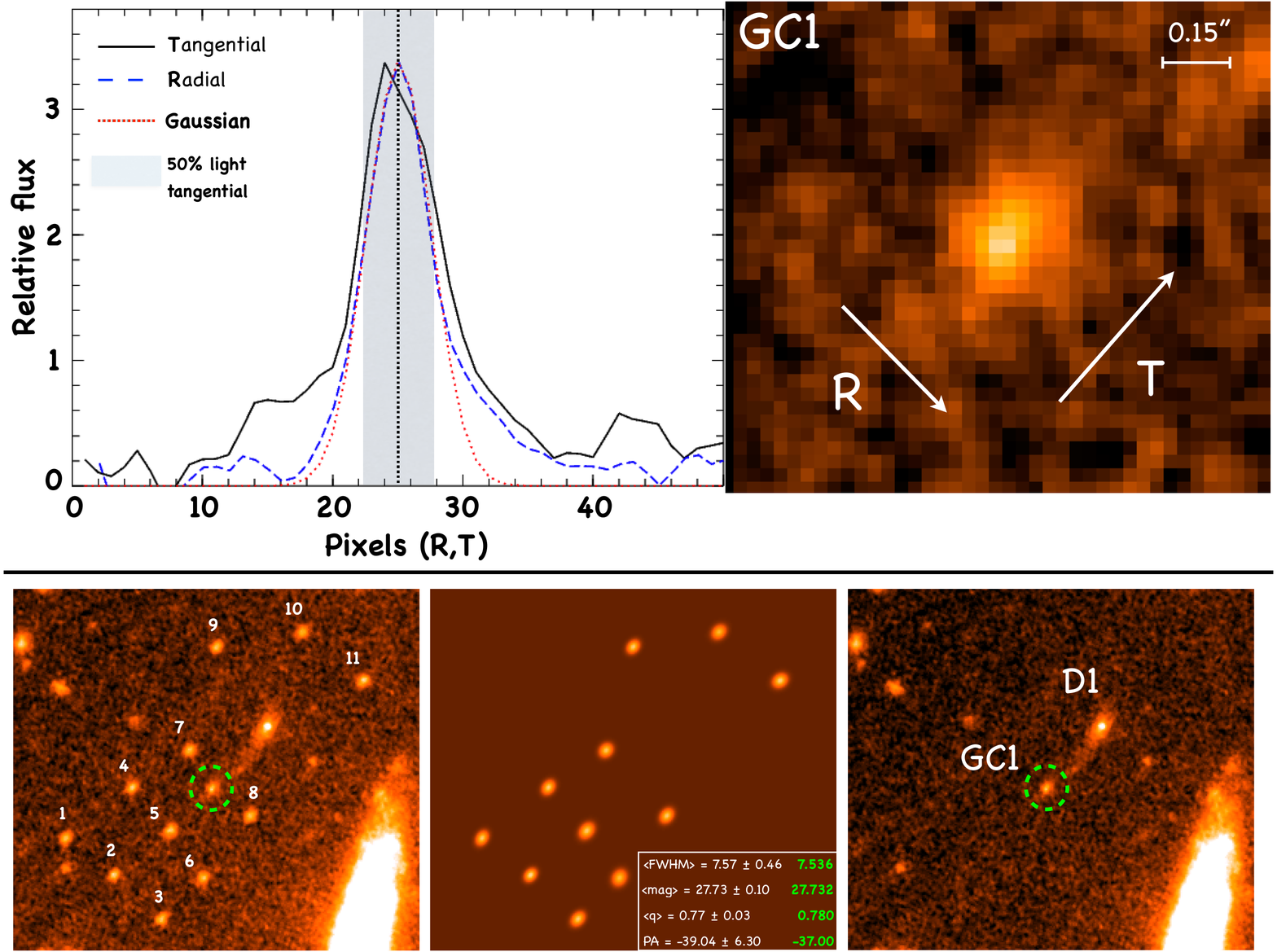}
\caption{Top left:  light profiles of the source GC1, at
  $z=6.145$, along the radial (R) and tangential (T) directions.  The
  range including 50\% of the light along T is marked with a gray
  region, while the profile is consistent with the PSF along R.  The
  radial and tangential directions are shown in the GC1 image to the
  right, where a slightly elongated shape is evident along T. 
  Bottom: eleven simulated images (1-11) inserted around GC1
  (dashed green circle) are shown (left); the {\tt Galfit} models and
  residuals are shown in the middle and right panels,
  respectively. The inset in the middle panel compares the average and
  standard deviation of the parameters recovered with {\tt Galfit} 
  (in white) with the real input values (in green).}
\label{GC1}
\end{figure*}

The observed spatial tangential profile of GC1 is shown in Figure~\ref{GC1}, in which the 50\% of the
light is enclosed within $\simeq5.5-6$ pixels as obtained with {\tt Galfit} taking into account the PSF. 
We explored a large grid of the most relevant parameters, the S\'ersic index, 
effective radius, total magnitude, the location of the source, the position angle and the axis ratio ($q=a/b$). 
To accomplish this task we have followed two different routes: 
(1) by allowing {\tt Galfit} to minimize its internal merit function and (2) following the method described \citet{vanz15}, 
by running {\tt Galfit} on a large grid of (fixed) values and monitoring the residuals of the ``observed-model'' image,
step by step.
While the light profile is not reliably constrained (e.g., we obtain a
good fit with both Gaussian and S\'ersic $n=4$ profiles, see
Figure~\ref{galfit}), the size in all cases is relatively well
constrained with $R_e$ not greater than 3 pixels (at
most). Specifically, the best estimates obtained in the case of a
Gaussian ($n=0.5$) profile is $R_e=1.4$ pixels.  An estimate of the
uncertainties has been obtained by inserting simulated images with sizes
and magnitudes similar to those of GC1 (Figure~\ref{GC1}), including
also the local noise and background gradients due to the galaxy
cluster, and analyzed with {\tt Galfit} following the same procedure
used for GC1. All the parameters are well recovered on average, with a
scatter that provides the typical statistical error (the results are
reported in Figure~\ref{GC1}).  We conclude that the error associated
to the observed effective radius of GC1 is of the order of half a
pixel in this specific case. Systematic uncertainties, however, could
dominate the error budget, in particular the unknown light profile and
total magnitude. For example, we allowed the source to be 0.4 magnitudes
brighter (27.50) than the measured F105W flux ($27.88\pm0.09$) and
explored S\'ersic profiles up to $n=10$. The effective radii are 0.9,
1.4 and 2.1 pixels in the case of magnitude 27.50 and n=8, 4 and 0.5
(Gaussian), respectively, while it is smaller than 1.2 pixels in all
the cases with magnitude 27.80 (the observed value).  Examples of {\tt
  Galfit} models are shown in Figure~\ref{galfit}. We conclude that a
plausible estimate of the radius is $R_{e}=1.7\pm 0.7$ pixels
(Figure~\ref{GC1}).

At $z=6.145$, with the tangential magnification computed above,
$\mu_{T}=17.5$, the inferred radius is $R_e=16\pm7$ pc.  The
circularized radius is therefore $R_c \simeq 20$pc, adopting $q=0.15$.

Even considering conservative estimates for the radius and magnification (i.e., $R_e=2.4$ pixels and $\mu_{T}>10$) the size
still remains remarkably small, $R_e<40$pc. 

\subsubsection{The system D1 and additional extremely faint SF knots}

\begin{figure*}
\centering
\includegraphics[width=17cm]{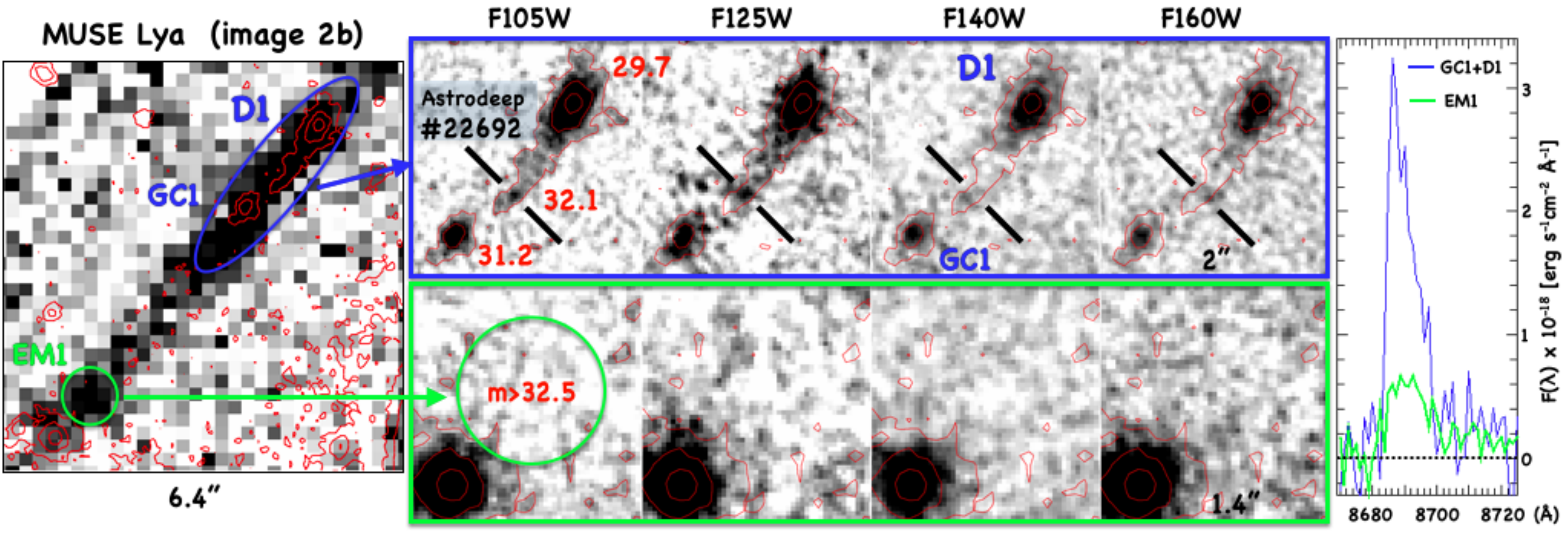}
\caption{A zoomed view of the Ly$\alpha$ emission (MUSE) of GC1 and D1
  (marked with a blue ellipse) at $z=6.145$.  The cutouts are shown in
  the F105W, F125W, F140W and F160W bands at the HFF depth (magnitude
  limit 29.4 at 3-$\sigma$ within 0.4$''$ diameter aperture).  The red
  contours show 2, 4 and 10-sigma level above the background.  D1 shows
  a clear nucleated core and an elongated emission along the
  tangential direction, suggesting it is compact in the inner part with
  an underlying distorted shape (modulated by the magnification).  The
  10-$\sigma$ contour of the core of D1 appears symmetric despite the
  large tangential magnification, suggesting it is significantly
  compact.  A possible stellar stream linking D1 and GC1 is present,
  traced by the 2-sigma contour.  Another knot, marked with ID
  \#22692 (Astrodeep, \citealt{castellano16b,merlin16}), is
  detected in the F105W and F125W bands, and barely in the F160W (top
  panels) suggesting a steep ultraviolet slope.  A Ly$\alpha$ emission
  feature without any HST counterparts is shown in the bottom panels
  (EM1, green circle).  The Ly$\alpha$ profiles for both GC1 + D1
  sources (blue line) and EM1 (green line) are shown to the right.
  The observed line fluxes are ($3.0 \pm 0.1$)$\times 10^{-17}$
  within a polygonal aperture and ($0.6 \pm
  0.2$)$\times 10^{-17}$cgs within a circular aperture of 1.6$''$
  diameter.  The de-lensed magnitudes of each object are shown in
  red in the F105W images.}
\label{D1}
\end{figure*}

The same {\tt Galfit} fitting has been performed for D1, the most
extended object among those studied in this work and close to GC1,
both at $z=6.145$.  A clear nucleated star-forming region surrounded
by a diffuse emission extending approximately $\simeq 200$ pc along
the tangential component is evident (Figure~\ref{D1}).  We obtain an
effective radius $R_e\simeq140$ pc, making this object compatible with
a forming dwarf galaxy. A morphological decomposition is beyond the
scope of the present work, but it is worth noting that the spatial
distribution of the nuclear emission is quite symmetric despite the
large magnification along the tangential direction (see the 10-sigma
contours in Figure~\ref{D1}).  As discussed above, such a symmetric
shape implies that the size of the inner part is extremely compact
also for D1.  The radius of the region enclosed within 10-sigma from
the background is $\sim 0.12'' \simeq 50$ pc 
suggesting a dense and very nucleated star formation activity.

Looking carefully at the image 2b, we identify even fainter features.
Figure~\ref{D1} shows the identification of an additional knot between
D1 and GC1, identified as ID=22692 in the Astrodeep catalog. Despite
the strong lensing magnification, its observed magnitude of F105W
$\simeq28.5$, implies a de-lensed F105W$\simeq32$ (i.e.,
$M_{UV}=-14.7$, adopting the aforementioned total magnification, 25).
From the SED fitting we derive an intrinsic stellar mass of $M \simeq
0.5-0.6 \times 10^{6} M_{\odot}$ and a $SFR \simeq
0.06$\msunyr. Although these measurements are still uncertain, lensed
sub-structures like this one provide a first glimpse into a completely
unexplored luminosity, mass domain at these redshifts.  The 
{\tt  Galfit} fitting does not provide in this case robust results, though
this object appears extremely small, at the level of a few parsec.
Understanding the nature of forming knots of this
kind will require JWST observations.  We also note that another
Ly$\alpha$ emitting knot is detected in the MUSE data, denoted as
``EM1'' in Figure~\ref{D1}, which does not have any counterpart in the HST
images, down to magnitude limits of 29.4 (at 3-sigma level, the HFF
depth). If this Ly$\alpha$ emission is produced by an underlying
star-formation activity it would imply that the associated source is
fainter than magnitude $\simeq33$ (i.e., fainter than $M_{UV}=-13.7$)
and the resulting rest-frame Ly$\alpha$ equivalent width larger than
300\AA. Alternatively, the Ly$\alpha$ emission may arise from
fluorescence induced by, e.g., GC1 and/or D1, suggesting 
the presence of surrounding neutral gas illuminated by close
star-forming activity.


\subsection{The source D2 in MACS~J0416 at $z=6.145$}
From the MUSE data-cube, we identified another lensed Ly$\alpha$
emission line of an object (named D2) at the same redshift as system 2
($z=6.145$), which is not part of the same galaxy (see
Figure~\ref{arc6}) and has a rest-frame Ly$\alpha$ equivalent width of
140\AA.  The optical counterpart is well detected in the HFF deep
photometry with a F105W magnitude of $28.33\pm 0.09$
\citep{castellano16b,merlin16}.  The object is located at $\simeq 27$
kpc from GC1 in the source plane, and is therefore distinct from
system 2 but plausibly part of the same environment of GC1 and D1.
The source is well fitted with a two-dimensional Gaussian shape and a
S\'ersic $n=4$ profile, with errors on the morphological parameters
dominated by the relatively low S/N. In practice, the object is not
spatially resolved, therefore only an upper limit on the effective
radius can be obtained. Using {\tt Galfit} on a grid of parameters and
simulations, as previously done for GC1,we can exclude an effective
radius greater than 1.5 pixels.
Therefore, adopting $R_e<1.5$ pixels and  $\mu_{tot} \simeq
3.0 \pm 0.5$ (in this case $\mu_{T} \simeq \mu_{R}$), we obatin an
intrinsic size of $R_e \simeq R_c < 150$ pc.

  \begin{figure*}
\centering
\includegraphics[width=17cm]{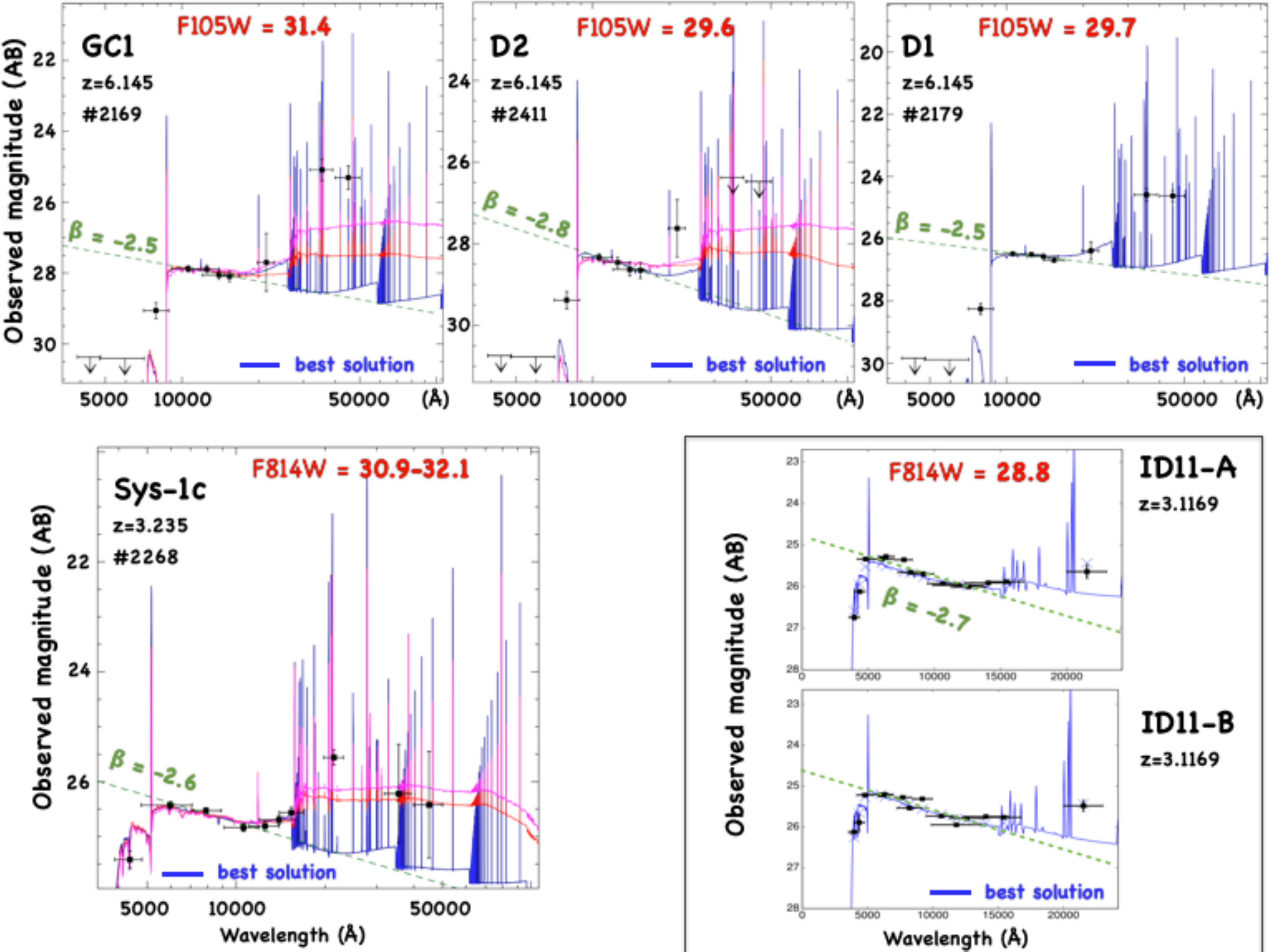}
\caption{SED fitting for each system in MACS~J0416, based on the Astrodeep photometric catalog \citep{castellano16b,merlin16}. 
The physical quantities derived from SED fitting of ID11 have already been discussed in \citet{vanz16a} and updated here with the deepest HFF photometry. 
The photometric redshift and the stellar emission including nebular prescription \citep{castellano16b} reproduce well
the observed magnitudes. The physical properties are summarized in Table~\ref{infos}. Best-fit solutions are shown in blue (see Table~\ref{infos})
and when present, old and more massive solutions are reported with red and magenta lines (300 and 700 Myr old, respectively).
The ultraviolet slopes $\beta$ are shown along dashed green lines.
De-lensed F105W and F814W magnitudes are shown in red in each panel.}
\label{SEDs}
\end{figure*}

\section{Physical properties from  SED fitting}\label{sedfitting}
\begin{figure}
\centering
\includegraphics[width=9cm]{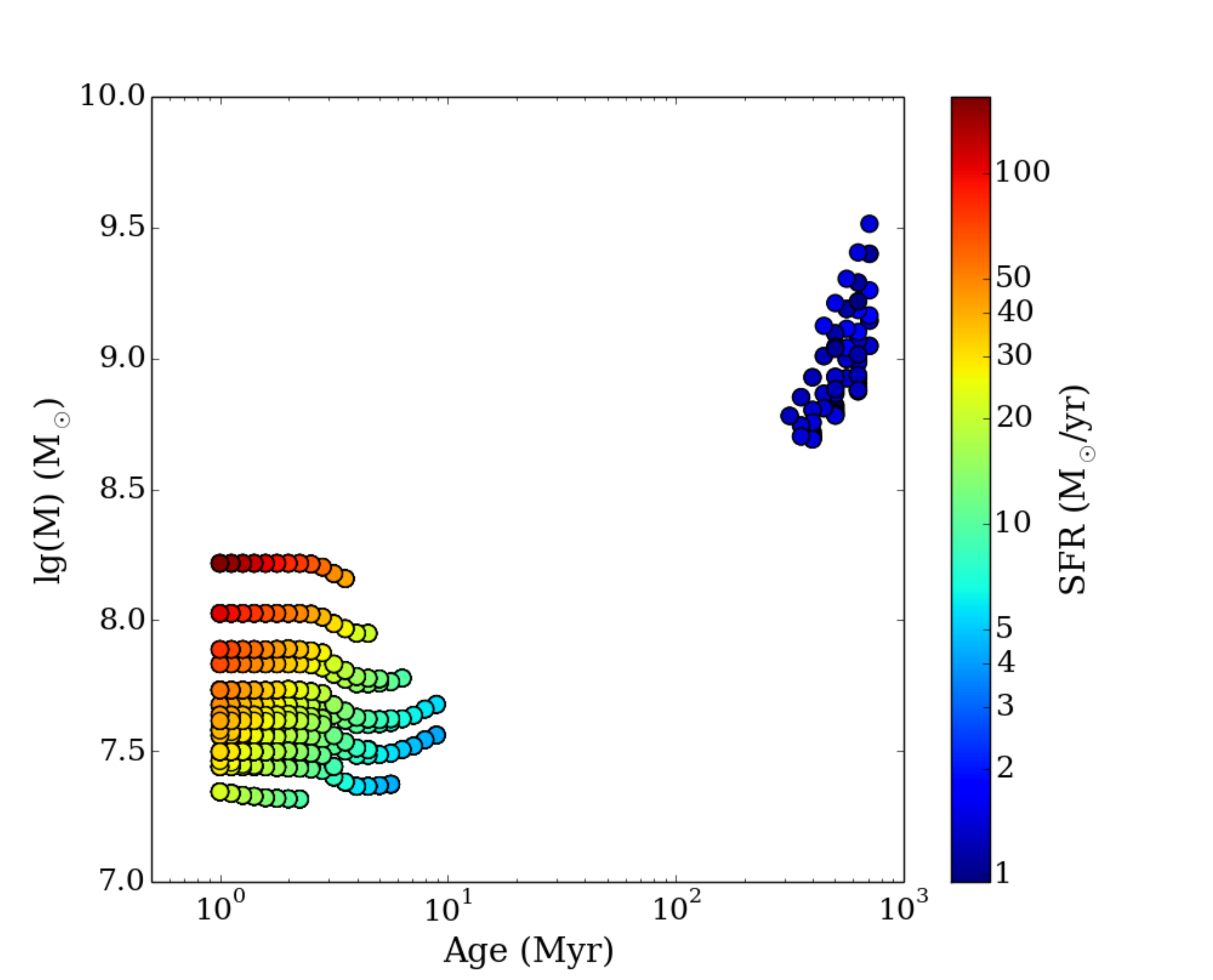}
\caption{The degeneracy among stellar mass, age and SFR
is shown in the case of the SED-fitting of GC1. 
All the solutions within the  68\%  interval are shown. The reported quantities are
derived from the observed photometry, i.e., they are not de-lensed (intrinsic values of the SFR
and stellar mass are obtained dividing by $\mu_{tot}=25$, see Table~\ref{infos}).
The SFR is color-coded and indicated on the right of the figure. 
The minimum SFR inferred from the \lya\ emission ($>4$\msunyr) 
favors the solutions at low masses ($10^{7.3-8.3}$\msun\ observed,
corresponding to $10^{5.9-6.9}$ \msun, de-lensed) and young ages, 
$1-10$ Myr. See the best solutions in Figure~\ref{SEDs}.
}
\label{degene}
\end{figure}

Physical properties of the aforementioned sources have been derived
from their spectral energy distributions (SED) by means of fits
performed with a set of templates from
\citet{BC03}, 
with the addition of nebular continuum and emission lines as described
in \citet{castellano16b} (see also \citealt{schaerer09}).  All the objects lensed by the HFF cluster MACS
J0416 benefit from the Astrodeep photometric catalog \citep{merlin16}
and the redshift measurements from MUSE observations. 
SED-fitting of the Astrodeep sources in MACS~J0416 was presented in
\citet{castellano16b} on the basis of photometric redshifts: here we
update that analysis by fixing the redshift at the spectroscopic value
and allowing also for very young ages (1-10 Myrs) that were not previously
considered.  The source ID11, behind the galaxy cluster
AS1063, has already been analyzed and described in \citet{vanz16a}.
Here we add the deepest near-infrared photometry that was not
available at that time. The resulting SED-fitting for all our objects
are shown in Figure~\ref{SEDs}.  The relevant parameters are reported
in Table~\ref{infos}, i.e. stellar masses, ages, E(B-V), star
formation rates along with their 68\% c.l.  uncertainties. The
observed photometry and SEDs are shown in
Fig.~\ref{SEDs}.  
It is worth stressing that, despite the extremely faint intrinsic
magnitudes involved (between $28.6$ and $31.4$), the multi-band
photometry is robust thanks to long HST exposure times in addition to
strong lensing magnification. For  the whole sample, 
typical S/N ranges from 10 to $>50$ in the ultraviolet continuum and
rest-frame optical wavelengths (for ID11 and Sys\_1c).  From a linear
fit of the observed photometry in the UV rest-frame range \citep[see
e.g.][]{castellano12}, we measure very blue ultraviolet slopes, ranging
between $-2.5$ and $-3.0$, which reflect the typically young ages,
very low dust extinction and possibly low metallicities
\citep{castellano14}.  In general, the intrinsic faintness of these sources
translates into modest star formation rates, of the order of $0.1 - 1$
\msun\ and low stellar masses, ranging between $1-20 \times 10^{6}
M_{\odot}$.  Such low masses have also been  measured by
\citet{karman16} behind AS1063 (including ID11). In
the case of GC1, D2 and Sys\_1c, solutions at higher masses and old
ages (>100Myr) are allowed by the fit due to larger uncertainties
in the Ks and IRAC fluxes. This has been verified by applying the photometric
errors of GC1 to the D1 object. This affects mainly the optical rest-frame part of the
SED, the Ks and IRAC bands. The solutions from the SED fitting of D1 show
similar degeneracies as observed for GC1, confirming that optical rest-frame photometry is
critical when inferring the ages and stellar masses. 
Older and more massive solutions are reported in Figure~\ref{SEDs} 
(red and magenta lines), in addition to the best fit
solutions (blue lines). However, the
Ly$\alpha$ flux measured from MUSE spectroscopy allows us to strengthen the
constraints on physical parameters for the two systems at
$z=6.145$. Following \citet{schaerer03} and assuming Salpeter IMF and
metallicity $Z=0.004$, the observed Ly$\alpha$ luminosity of GC1 and D2
can be converted into a lower limit\footnote{The Ly$\alpha$ emission can be partially 
attenuated by a small amount of dust and by the presence of the intergalactic medium.} 
of SFR = 4 \msunyr\ and 2 \msunyr, respectively.  
With these additional constraints, no solutions older than
9 Myrs are found for GC1 such that the upper limit on its stellar mass
is reduced by a factor $\sim$20 from 3.3 $\times 10^{9}$ to 1.6
$\times 10^{8}~ M_{\odot}$ (observed). Similarly, the maximum age allowed for D2
decreases from 710 to 100 Myrs with a maximum stellar mass of 2.3
$\times 10^{8}~ M_{\odot}$ (observed).  An example of the aforementioned degeneracy 
among stellar mass, age and star-formation rate is shown in 
Figure~\ref{degene} for GC1, the smallest of our objects. A main ``cloud'' of solutions
is visible and spans the intervals $1-10$ Myr and  $10^{7.3-8.3}$ \msun\ (observed).
A second ``cloud'' at high masses and older ages is also apparent, however it is 
 disfavoured if the star-formation rate derived from the \lya\ emission is considered 
(SFR(\lya) $>4$ \msunyr).
It is also worth stressing that solutions with larger masses would also imply stellar mass
densities more than 10 times larger than the best-fit values reported in Table~\ref{infos}.
Therefore our objects are fully consistent with
solutions favoring very young ages, low stellar masses and low amount of dust.
The combination of the sizes and the physical quantities described above
allow us to estimate the surface densities of star-formation and
stellar mass for the objects of our sample.
A consequence of the inferred stellar mass and star formation rate estimates is the very large 
specific star formation rate (sSFR=SFR/M$_{\star}$), a quantity that does not depend on the magnification.  
Specifically, we derive sSFR spanning the interval 30-800 Gyr$^{-1}$, as expected for
young, low mass object in which a burst of star formation is ongoing \citep[e.g.,][]{karman16}.
In particular, the object GC1 shows a sSFR $\simeq 800$ Gyr$^{-1}$ and will double its stellar
mass in $\lesssim 5$ Myr.

Overall, it is also worth noting that a significant contribution from the nebular emission lines
\oiiidoub, \hb\ and \ha\  is expected in the Ks or IRAC bands, at the level of several hundreds or
thousands \AA\ of equivalent width (rest-frame). In particular, the
nebular contribution in the Ks-band for ID11 has been well measured with
VLT/X-Shooter, allowing us to derive robust estimates of its physical properties from
SED fitting \citep[][]{vanz16a}. Similarly, X-Shooter observations of Sys\_1 at $z=3.235$
are under acquisition. For the objects at $z>6$ presented in this work there is a clear signature
of nebular line contribution (\oiiidoub,  \hb\ and \ha) in the IRAC $3.6\mu m$ and $4.5\mu m$ channels.
The precise intensity and line ratios at $z>6$
will only be measurable in the future when JWST NIRSpec spectroscopy will become available.
The rest-frame optical stellar continuum will also be observable with JWST NIRCam. 

\section{Discussion}

\subsection{Proto-Globular Clusters}

%

Current scenarios for the formation of
multi-population GCs include a massive first generation (FG) of stars
forming at redshift $z>2$ \cite[e.g.,][]{kruijssen15, renzini15}.
At the present time, little is known about the
physical conditions characterising the stellar FG.  From the
population of GC present in the Milky Way, we know that in general and
in terms of mass, present-day GCs have sub-dominant FG populations
with respect to second generation (SG), with a very few exceptions
\citep{bragaglia15}.  Current theoretical multi-population scenarios
for GC formation do not focus much on the events originating the FG.
The most popular scenarios, i.e. the AGB and the fast rotating massive
stars scenarios, generally start keeping track of the evolution of the
system after a FG is already in place, generated by a single,
instantaneous burst of star formation occurred at high redshift
\cite[$z>2$, e.g.,][]{renzini15,kruijssen15}.  In both scenarios the
SG of stars forms out of the gas shed by FG stars. As the mass return
from aging stellar populations is in general too scarce to form a
large SG population \cite[assuming a standard IMF;
e.g.,][]{calu14,renzini15}, in order to account for the present-day
mass and predominance of SG stars as observed in GCs, the FG has to be
substantially more massive than present-day GCs, by a factor ranging
between 5 and 20 \citep{dercole08, renzini15}. Most of this massive FG
has then to be lost via dynamical processes, such as evaporation and
tidal interactions.
Regarding this aspect, it is noteworthy that a system like CG1 presents
hints for a dynamical interaction with the massive companion D1,  
and for a possible stellar stream between these two objects. 
This could indicate an already on-going stellar stripping, as expected in
GC scenarios soon after the formation of the FG \cite[e.g.,][]{dercole08}.   

The fast-rotating massive stars (FRMS) scenario \citep{decressin07} assumes 
an initially highly concentrated cluster with a small half-mass radius, typically of the order of a few pc  \citep{decressin10, krause12}, 
with initial total mass densities comparable to the central densities of the most massive clusters observed today ($\sim 10^5$ M$_\odot$ pc$^{-3}$, \citealt{renzini15}). 
The expulsion of the residual gas is assumed to occur on a relatively short time-scale, faster than the crossing timescale \citep{decressin10}, 
soon after the cease of type II SN explosion as due to a sudden accretion onto the dark remnants  \citep{krause12}. 
The loss of the gas causes a sudden change of gravitational potential, which is then able to unbind a large fraction of first generation low-mass stars sitting initially
in the GC outskirts \citep{decressin10}. 

The AGB scenario assumes a GC which forms within the disc of a high-redshift galaxy \citep{dercole08, dercole16}. 
A natural outcome of the AGB scenario is a FG characterised by a flatter density profile than the second generation. 
This is due to the fact that after the explosion of all the SNe of the FG, the gas shed by the AGB stars originates a 
cooling flow directed towards the centre of the cluster. 
As the SG forms out of this gas mixed with some pristine gas \citep{dercole16}, its distribution will be much more concentrated 
than the one characterising the FG \citep{dercole08}. 


A key element which differentiates the AGB and the FRMS scenarios is the size assumed for the stellar FG. 
The initial size and concentration of the FG is a crucial quantity regulating the subsequent mass loss experienced by the 
cluster in the remainder of its history.
\citet{dercole08} study two different models characterised by a FG of mass $10^7$ and $10^6$ M$_{\odot}$, 
which correspond to GCs of present-day masses of 
$10^6$ and $10^5$ M$_{\odot}$, respectively. In both cases, the distribution 
of the FG follows a \citet{king62} radial profile, with a 
half-mass radius of $35$ pc and $16$ pc for an initial mass of $10^7$ and $10^6$ M$_{\odot}$, respectively. 
For each model, the truncation radius of the profile corresponds to the  
tidal radius at a distance of 4 kpc from the Galactic Centre. 
This results from the assumption that the GC is placed on a circular orbit located at this galactocentric distance and  
with an external, galactic tidal field modelled as a Keplerian potential generated by a point mass of 
$M_{g} \sim 4 \times 10^{10}$ M$_{\odot}$. 
These assumptions ensure a significant mass loss of FG stars due to the external potential. 
If the energy injected by the FG stellar winds and SNe is sufficient to expel 
the SN ejecta and the residual gas \citep{calu15}, the stellar FG can expand beyond its tidal 
limit in response to this substantial gas loss and be prone to efficient stellar mass loss due to the external field \citep{dercole08}. 
Clearly, the efficiency of this mechanism is sensitive to the parameters 
regulating the initial FG distribution, 
as more concentrated stellar distributions will give place to smaller amounts of mass 
lost via tidal stripping \cite[e.g.,][]{vesperini97}.

Beside supporting a preferential loss of FG stars, the natural prediction of the AGB scenario of a 
a more concentrated SG is also in agreement with observations of GCs, showing that red stars (generally identified with SG) are
always more centrally concentrated than blue stars (generally identifiend with FG stars; \citealt{lardo11}).  
These aspects outline another key difference between the AGB and FRMS scenario, i.e. that in order to explain the 
different proportions and radial distributions of FG and SG stars, the latter has to postulate that the SG was formed close to the massive
stars in a mass-segregated star cluster \citep{decressin10}.

The ab-initio study of GC formation of \citet{nakasato00} starts from 
a metal-free proto-GC cloud of a few $\sim 100$ pc, in which a first population of $10^2$ M$_{\odot}$ of 
metal-free stars originate, enriching the cloud with heavy elements and whose 
self-generated emission of photo-dissociative photons stops the early, nearly instantaneous burst. 
In the polluted gas, density perturbations are allowed to grow via thermal instability (see also \citealt{fall85}), 
and radiative processes lead to a strong condensation in the cloud which causes a burst of star formation, which 
in $\sim 10$ Myr produces a stellar population of $10^5$ M$_{\odot}$. The stellar mass profile 
calculated at  $\sim 10$ Myr is characterised by a half-mass radius of $\sim 30$ pc, i.e. consistent with the values
assumed in the AGB scenario. \\
The compact systems GC1 and ID11 present stellar densities, stellar masses and half-light radii consistent with the FG stellar masses 
and half-mass radii expected from the AGB scenario, and radii also compatible with the expectations of \citet{nakasato00}. 

It is worth stressing that the half-light radii are determined from 2D light distributions 
and should be regarded as underestimates of the 3D half-mass radii (typically of a factor $\sim 3/4$ for a S\'ersic profile), 
with a weak dependence on the S\'ersic index, see \cite{wolf10}.
Our study shows the existence of very compact stellar objects in a redshift range in which GCs should be actively forming.
Beside the sizes, in at least two cases (GC1 and ID11), also the ages of the stars are compatible with those expected for 
GCs caught  during the formation of their stellar FG. 

\begin{figure}
\centering
\includegraphics[width=8.5cm]{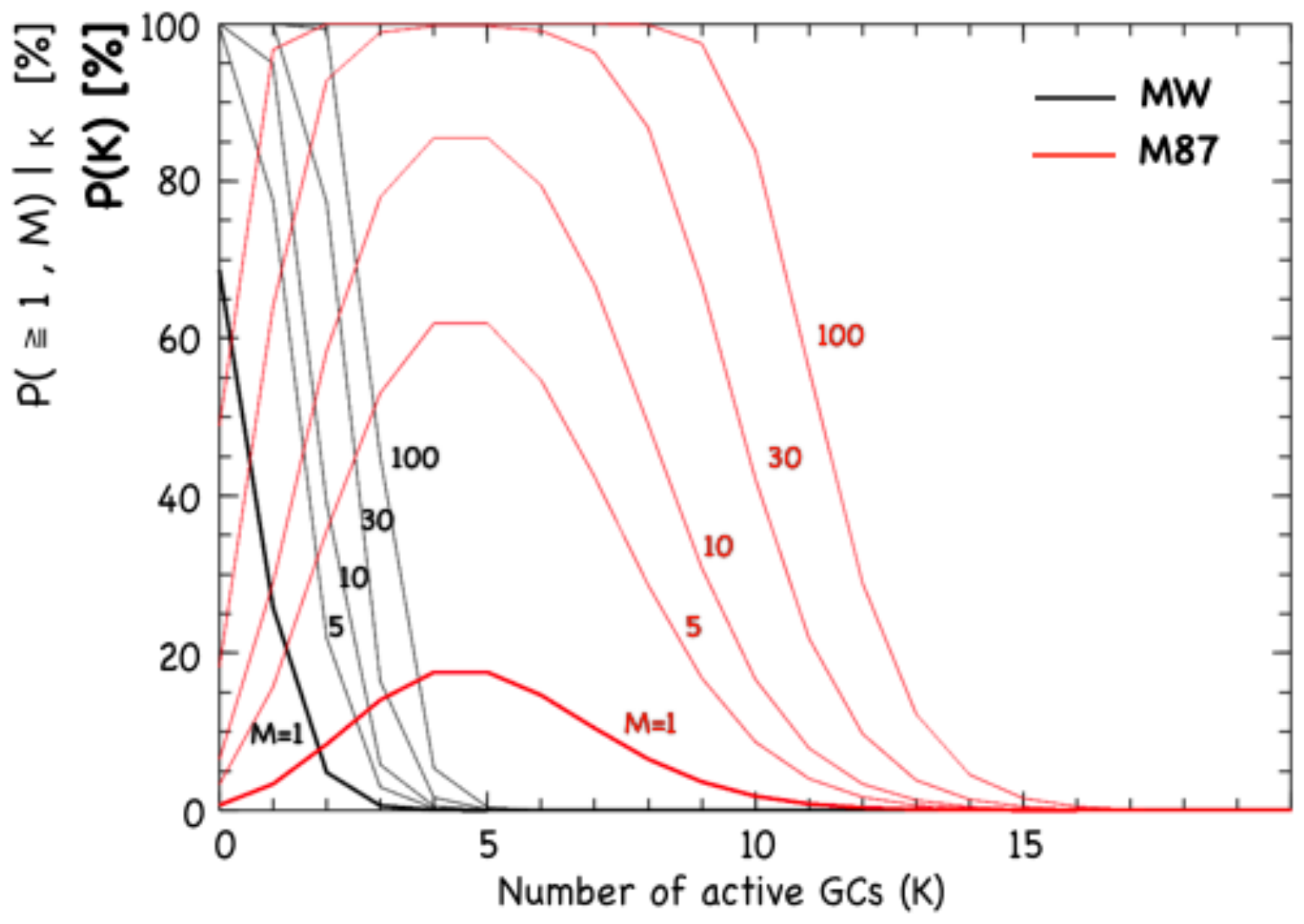}
\caption{
Two plots are shown in the same panel: the thick label in the Y-axis corresponds to the thick lines
and the thin label to the thin curves. In particular,
the thick black and red lines represent the probability $P(K)$ that K GCs are simultaneously active
(i.e. in a star-forming phase) at $z>3$ in a MW-like and in  a M87-like galaxy, respectively. 
For example the probability $P(K)$ that 1(5) GCs are forming in the MW-like or M87-like galaxies
is 26\% and 18\%, respectively.
The thin black and red lines are instead the probability $(P\ge1,M)|_{K}$ that K simultaneously active GCs are
present in at least one galaxy in a sample of M MW-like and M87-like galaxies, respectively. 
The quantity $(P\ge1,M)|_{K}$ is computed for M=1, 5 ,10, 30 and 100. 
For example the probability that 5 (=K) GCs are active in one galaxy out of a sample of
10 (=M) observed M87-like galaxies ($P\ge1,M)|_{K}$ is 85\%.
By definition $\int_{0}^{\infty} $P(K)$ dK =1$.}
\label{PK}
\end{figure}

\subsection{The chance to observe a forming GC}
\label{sec_chance}

It may be useful to determine what is the probability to observe a GC in its forming phase (hereafter {\it active} GC).
The physics of GC formation is quite uncertain \cite[e.g.,][]{renzini15}; in this work, we 
adopt a conservative approach by making plausible assumptions. 
To this aim, we consider two cases studied in the local Universe: the Milky-Way and the giant
elliptical galaxy M87. This choice is due to the fact that 
MW-like mass star-forming galaxies might be rather common at high redshift and visible as Ly$\alpha$ emitters 
\cite[see e.g.,][]{dayal12}. 
On the other hand, owing to its large stellar mass ($\sim 7 \times 10^{11}$ M$_\odot$, \citealt{forte12}) and 
to the presence of thousands of GCs \citep{bellini15} 
M87 should be regarded as an extreme case and at high redshift it will be much rarer than MW-like mass 
galaxies. 

It is known that the MW is surrounded by $N \simeq 150$ GCs \citep{harris96},
whereas \citet{bellini15} have identified almost $N \simeq2000$ GCs in the core of M87. 
In our calculation we assume that 
in a generic GC the first star formation event lasts $\Delta T_{GC}=5$Myrs. 
This time scale corresponds to the typical age of local young massive clusters, 
which are known to be able to retain their gas for only a few Myr after the formation of their stars 
\cite[e.g.,][]{bastian13}, either because of a particularly 
strong feedback favouring gas expulsion, or because at  least all the strongly
gravitationally bound gas is  converted into  stars \citep{charbonnel14}. 

The second assumption is that a GC formed
sometime within the first two Gyrs after the Big-Bang, 
corresponding to $z>3$ or $\Delta T_{epoch} \simeq 2$ Gyr.  

We make the further conservative assumption that the formation epoch follows 
a flat distribution. Clearly, the assumption of a formation epoch peaking at some particular cosmic time 
would increase significantly the probability of catching an active GC around that cosmic time. 
 
As we also ignore the original spatial distribution of proto-GCs,  
we assume that all GCs formed in the vicinity of the dark matter halo hosting
the main galaxy, such that a typical observed field of view probes a sufficiently large volume to spatially include 
all the proto-GCs. In other words, we neglect if a GC has been acquired or formed ``in-situ'' \citep{katz14}.
This assumption is corroborated by cosmological simulations indicating that proto-GCs 
from within a distance of tens of kpc proper from the main dark matter halo in which they are embedded, 
a volume well sampled by the observations  \cite[e.g.,][]{kravtsov05, ricotti16}.

Under these simple hypotheses, the probability (P) to observe at a given cosmic time (at $z>3$), 
$K$ (or $\ge K$) active GCs in a sample of $N$ ($>K$) GCs is: 

\begin{equation}
P(K)=\left(^{N}_{K}\right)p^{K}(1-p)^{N-K};~~~~~~P(\ge K)= \sum_{i=k}^N P(i);
\label{eq1}
\end{equation}


\noindent where the probability $p$ of the single case is p=$\Delta T_{GC}$  /$\Delta T_{epoch}$=0.0025 (as mentioned above
by assuming a flat distribution). The probability that {\it at least} 1(5) out of $N$ GCs is caught during their formation
is $\sim30(56)$\% in the case of MW(M87), sometime at $z>3$. Clearly these probabilities increase (decreases)
if $\Delta T_{GC}$ decreases (increases) or $N$ increases (decreases).
It is worth stressing that if a non-flat formation history  was assumed for proto-GCs, then we would have 
a specific cosmic epoch in which the frequency of simultaneously active GCs would be particularly high. 
In this respect, the probabilities computed with the assumptions described above should be regarded as conservative. 

If we perform the same exercise by assuming a sample of  $M$ galaxies,
the probability will obviously increase. The probability that an event  (with probability P(K)) occurs in at least one of the observed galaxies 
is $1-[1-P(K)]^{M}$.

The probabilities that 1, 2, or 3 GCs are active in a MW-like galaxy are $P(K)$=25.8,4.8, and 0.6\%, respectively.
Moreover, the probabilities to observe 1, 2, or 3 active GCs in {\it at least} one  out of ten (M=10) MW-like galaxies at $z>3$ 
are 84.6\%, 16.4\%, and 1.2\%. 
In the case of M87-like galaxies these numbers increase significantly, although such massive objects are rarer. 
Figure~\ref{PK} shows the probability $P(K)$ as a function of $K$ calculated for various values of the sample size $M$
under the aforementioned hypotheses.  In the same figure, the black and red thick lines show the probabilities
P(K) as calculated from Eq.~\ref{eq1} for MW and M87, respectively, whose underlying area is one by definition.
The thin lines show the probability that  in {\it at least} one galaxy out of M  there are K active GCs,
for a given $P(K)$ ($P\ge1,M|_{K}$).

At the current stage, it is very difficult to compare the quantities shown in Fig.~\ref{PK} with 
any of the observable properties discussed in the current work. 
To perform such a task, calculations of simultenously active GCs per unit volume would be 
required, where number densities need to be extracted from cosmologically-based models computing detailed merging trees, 
as performed e.g., in \citet{ricotti02}. The calculation described in this section shows 
that current surveys of lensed fields likely contain many active GCs, and that 
the probability to observe a few of them simultaneously in a star-forming, active phase is quite high. 
A more detailed comparison between expected frequency and observed number of star-forming proto-GCs is postponed to the future, 
when larger samples of similar objects will become available.
In the next section the observability of such faint objects is discussed.  

\subsection{Can we observe active GCs?}

Despite the current GC formation scenarios are still uncertain \cite[e.g.,][]{renzini15}, 
we report in Table~\ref{infos} the possible ranges of a few relevant quantities 
expected during the formation of the first generation stars in GCs of two different masses and within the AGB scenario 
\citep{dercole08,dercole16}. 
\citet{dercole08} study the formation of one proto-GC of  stellar mass $10^{6}$ \msun\ and half-mass radius of 16 pc and 
another more massive one, characterised by a stellar mass $10^{7}$ \msun\ and an half mass radius of 35 pc.
They assume a FG of stars already in place and focus mostly on the star formation history of SG stars. 
Assuming for the FG a constant SFR occurring on a timescale of 5 Myr as adopted in Sect.~\ref{sec_chance}, 
we obtain for the low mass and high mass GC SFR values of 0.2 and 2  \msunyr\, respectively. 
The two systems present stellar mass surface densities of 800 and 1700 M$_{\odot}$ pc$^{-2}$ for the less and the
more massive cases, respectively.
By means of the Starburst99 models \citep{leitherer14} we have
calculated the expected monochromatic luminosity at 1500\AA~in the
case of an instantaneous burst for the two stellar masses reported
above and at an age of $\sim 5$ Myr after the burst.  The result
depends weakly on the adopted initial mass function and metallicity.
The expected apparent magnitude at 1500\AA~rest-frame are 30.5 and
$\simeq 28$ at $z>3$ for the less massive and more massive cluster,
respectively.  This corresponds to absolute magnitudes fainter than
$M_{UV} =-17$ (consistently with estimates by \citealt{schaerer11}).  
Clearly these values are elusive in the deepest
non-lensed fields, such as the Hubble Ultra Deep Field (\citealt{beckwith06}), 
particularly at magnitudes fainter than 29.5.
However, as demonstrated in the previous sections, observations at the
HUDF-depth in regions of the sky magnified by lensing galaxy clusters
(HFFs) have allowed us to probe extremely compact stellar systems
($<100$ pc) at $z \simeq 3-6$. 

In addition, the expected Ly$\alpha$ line flux of proto-GCs is
consistent with the observed values.  The Ly$\alpha$ flux emerging
from a proto-GC can be estimated by assuming the case B recombination
and the SFRs reported in Table~\ref{infos}, yielding $0.5-3 \times
10^{-19}$\ergscm\ \citep{schaerer03}.  Ly$\alpha$ line emission at
these flux levels is in principle accessible  
in the deepest fields observed with MUSE (e.g., HDF-S,
\citealt{bacon15},  see also Figure~6 in \citealt{vanz16c}). However,
the detection of the continuum at magnitudes fainter than 30 is out of
reach or at the very limit (1-2 sigma detection) of the current
deepest field surveys.  This work demonstrates that strong
gravitational lensing in deep fields allow us to overcome these
limitations.\\
Finally, a previous study presenting considerable analogies with the current one and which is worth to mention is 
described in \cite{ellis01}. 
Also in that case, a blind spectroscopic mode search was performed of objects lying nearby the critical line of the 
lensing cluster Abell 2218. A multiply-lensed, intrinsically faint ($I \sim 30$), compact 
($<150~$ pc) system was found at $z\sim 5.6$ and with a very small stellar mass ($\sim 10^6~M_{\odot}$). 
A young age for such a system ($<2 Myr$) was inferred from an upper limit on the stellar continuum, with the SFR deduced from its 
Ly$\alpha$ emission. 
As the pioneering work of \cite{ellis01} shows several parallels with the one described here, it can 
be regarded as a remarkable forerunner of our results.

\subsection{Extremely faint, compact and dense forming objects}
The most compact objects, ID11 and
GC1, show physical properties that are not far from those expected for
proto-GCs.  Before discussing this topic, we briefly report on their
environment.

Object ID11 lives in a group of star-forming galaxies lying at the
same redshift as confirmed by MUSE \citep{karman16}, including a
Ly$\alpha$ nebula described by \citet{cam16b} located at $\simeq$
100 kpc from ID11.  Similarly, GC1 is part of a system
including D1 and D2 at the same redshift, $z=6.145$.  Also System 1 in
MACS~J0416 (at $z=3.235$) is possibly part of a group of galaxies at
similar redshift ($\Delta z < 0.05$), though in the current MUSE data
no other galaxies have been confirmed at the same redshift
\citep{cam16c}.  Therefore, our young, compact and low-mass objects
are possibly sharing the same environment of (slightly) more massive
and older systems. It is not clear if GC1 or ID11 will eventually
merge with other systems or will remain bound after the gas is removed
(i.e., will maintain their identity) until the present days.  This is
reminiscent of the present day old dwarfs and GCs (older than 10 Gyr)
surrounding the local galaxies (including the MW) and implying their
ancestors were rapidly forming stars on a short time-scale 
(as we are observing here).
Our objects might be the youngest counterparts of
the extremely old systems observed today.

It is worth stressing that the intrinsic (de-lensed) magnitudes of the
object discussed in this work span the range $29.0-31.4$. In
particular GC1 with its intrinsic magnitude of 31.4 is a clear example
of what is currently missed in the deepest surveys in the field, such as
the Hubble Ultra Deep Field (whose magnitude limit at 1-sigma is
$\simeq 30.5$). In this respect, the sources reported here open a new
window to a very low-mass/luminosity regime that unavoidably will need
JWST and subsequently the Extremely Large Telescopes for the
characterization of their physical properties, especially for what
concerns the absorption line science.

\subsubsection{Dynamical mass of ID11}
It is worth noting that ID11, the object with the best photometric
measurements ($S/N>30$) and with optical rest-frame spectroscopy,
shows extremely narrow emission lines, both the high ionization
ultraviolet lines (e.g., \civ, \heii, \oiiiuv, \ciiidoub) and the
optical lines (e.g., \oiiidoub). In particular, the oxygen (optical)
lines \oiiidoub\ are well detected ($S/N = 12-33$) and barely resolved
in our near-infrared X-Shooter spectrum ($R \simeq 5000$), implying a
very low velocity dispersion $\sigma_{v} \lesssim 20$ \kms\ (see
\citealt{vanz16a}), and also a low dynamical mass.  Following
\citet{maseda13}, we determine the dynamical mass using the virial
relation:

\begin{equation}
M_{dyn} = C \frac{R_e ~ \sigma_{v}^{2}}{G}
\label{eqdyn}
\end{equation}

\noindent where $R_e$ is the effective radius (assumed here to be the virial radius), 
$\sigma_{v}$ is the velocity dispersion 
\cite[e.g., ][]{rhoads14,maseda14}.  

In general, it is well known that the coefficient $C$ is weakly
dependent on the density profile when the velocity dispersion is
measuread over large apertures (in principle, over all the object),
and independent of orbital anisotropy \cite[e.g.,
see][]{ciotti91,ciotti94}.  In particular, for values of the S\'ersic
index of $n\simeq 4-5$ as found for this system, $C$ varies between
4.65 and 3.67 \citep{bertin02}.

As discussed in \citet{maseda13}, there are several potential
systematic effects that may affect these estimates, including the fact
that the measured half-light radius is not necessarily equal to the
virial radius and that the dynamical structure might be irregular and
not virialized.
ID11 does not show an irregular morphology
(plausibly close to spherical) and is well fitted by a single component profile, at least along 
the tangential direction. 

Assuming the above relation applies to ID11 and adopting $C=4$ as for the Green Pea galaxies \citep{maseda14, erb14}, 
the comparison with the stellar mass gives $M_{dyn} / M^{\star} \simeq 1$. Given the uncertainties in the
estimation of $R_e$ and the stellar mass, the ratio is fully consistent with a value of $\sim 1$.

Currently, there is no evidence that local GCs possess DM halos (\citealt{heggie96}, but see also \citealt{ibata13}). 
It is possible that GCs were originally embedded in DM halos which have then be stripped 
by the tidal field of the host galaxy \citep{mashchenko05}, or that DM may be still present in the outer regions of the
clusters  \citep{bekki12, ibata13}. 
Our results obtained for ID11 imply a dynamical mass dominated by the stellar mass.
Clearly, our result does not exclude the presence of DM on larger scales. 
In the future, it will be crucial to extend the study described in this Section to larger samples, possibly comparing the results obtained 
using different emission lines to estimate the velocity dispersion $\sigma_{v}$. 

\subsection{Low-mass objects as contributors to cosmic reionization}

\subsubsection{Low-mass object might have large Lyman continuum leakage: the need for rest-frame optical spectroscopy}

As recent hydrodynamical simulations of proto-GCs have shown \citep{calu15}, the feedback of the stellar winds and SNe belonging to the FG 
can produce large and elongated hot cavities along which their interstellar gas is able to escape. In principle, 
these channels may also represent viable escape routes for ionizing photons. However, in order to assess 
whether proto-GCs might be important reionizing sources, an estimate of the time span 
over which their ISM remains rarefied and ionized is required, as well as the covering factor of the 
hot cavities and its evolution with time. Currently, work is in progress to investigate these aspects. \\
On the observational side, the ionizing radiation emitted by faint objects as well as their opacity at the
Lyman continuum (LyC, $\lambda< 912 \AA$) can be investigated only by pushing observations towards lower and lower
luminosity regimes. While at relatively bright luminosities ($L>0.5L^{\star}$) 
the escape of ionizing radiation is not favoured, at least as far as observations at $z<4$ indicate 
  \cite[e.g.,][]{vanz16b, izotov16, shapley16} 
and in general very modest   \cite[e.g., $<1-10$\%,][]{vanz10, vanz12b, grazian16, guaita16, bridge10, siana10, siana15}, 
in the low-luminosity domain it is still poorly explored. 

Strong lensing magnification has
allowed us to detect very faint sources (de-lensed $m_{1500} \gtrsim 29-30$) observed at  $m_{1500} \gtrsim 26-27$ (implying magnification
factors  higher than 15).
In order to possibly detect the Lyman continuum emerging from the same objects, we would need observations $\gtrsim 3$ magnitudes 
deeper at 900\AA\ than at 1500\AA\ to probe an escaping radiation of 20\%, i.e. magnitudes of the order of 29-30 in the Lyman continuum 
are still needed even in strongly lensed fields  \cite[e.g.,][]{vanz12a}.

In this challenging scenario, and since during the reionization epoch ($z>6$) the Lyman continuum is not detectable owing
to cosmic opacity, we must rely on indirect probes of
Lyman continuum leakage, possibly calibrated on reference samples of Lyman continuum sources
at $z < 4$, when the IGM still allows us to directly detect the ionizing radiation. 
In this regard, recent progress has been made in the field of photo-ionization modeling  
\citep{jaskot13, nakajima14, zackrisson16, inoue11}, 
subsequently confirmed by observations \cite[e.g.,][]{debarros16, vanz16b, verhamme16}.
In particular, line ratios in the rest-frame optical band like the O32 index (\oiiiv/\oiilam) and specific properties
of the Ly$\alpha$ profile \citep{behrens14,verhamme15,dijkstra16} can provide valuable indirect probes of  
the physical state of the interstellar medium and of the column density of neutral gas (e.g., density-bounded or ionization-bounded), 
as well as the connection with the ultraviolet spectral slope, the Balmer emission lines and the Lyman continuum leakage
\citep{zackrisson13, zackrisson16}.

In \citet{karman16} a few lensed sources were already identified presenting a  low column density of neutral gas ($<10^{18}$ cm$^{-2}$),
possibly compatible with a Lyman continuum leakage (but still not confirmed directly). In particular, in the case of ID11, 
the VLT/X-Shooter near infrared spectroscopy will reveal remarkable properties never observed before in such a faint and distant 
object (see \citealt{vanz16a}). ID11 is a young, low-metallicity and dust-free object possibly captured during its first burst of star formation
and confined in a small volume, surrounded by a shell of expanding gas. 
It also shows a low column density of neutral gas ($<10^{18.5}$ cm$^{-2}$, though not necessarily optically thin to Lyman continuum)
and a large O32 index ($>10$) compatible with a density-bounded condition in the ISM.

Therefore, rest-frame optical spectroscopy is crucial in this line of
research.  VLT/X-Shooter observations for System 1 at $z=3.235$ are on
going (P.I. Vanzella) and will provide unique information at
rest-frame optical wavelengths and at twice the spectral resolution
of MUSE (as we did for ID11).  It will also be essential to improve the dynamical mass estimate.

At $z=6.145$, the access to rest-frame optical nebular lines like
\oiidoub, H$\beta$, \oiiidoub\ and H$\alpha$ (just to mention the most
relevant ones) requires the NIRSpec instrument on JWST, as well as
NIRCam for optical rest-frame morphology at 3-5$\mu m$. The comparison
of the ISM properties derived using the same spectral diagnostics
(i.e. at the same rest-frame wavelengths) for these low-mass and
extremely young star-forming objects will be crucial, especially in
cases where a leakage of ionizing radiation is confirmed for sources
at $z<4$.  This will eventually represent a unique training set for
the identification of the sources responsible for the reionization of
the universe.

\subsubsection{Ly$\alpha$ nebulae as possible indirect signature of multiple proto-GCs ionization power}

Using the \citep{schaerer02, schaerer03} stellar population models and assuming a constant star-formation
rate of 1~\msunyr, after $\sim 3$ Myr the expected Ly$\alpha$ luminosity
is $10^{42}$erg~s$^{-1}$. If we rescale this value to our SFRs (0.1~\msunyr), we obtain a Ly$\alpha$ luminosity
of the order of $10^{41}$erg~s$^{-1}$. 
As discussed earlier in Sect.~\ref{sec_chance}, the probability that at least 10 proto-GCs are active at the same time in M87-like
galaxies observed at $z>3$ is not negligible ($>10$\%). Therefore, assuming 10 proto-GCs simultaneously
in place and emitting Ly$\alpha$ photons,  their contribution to a diffuse Ly$\alpha$ emission might be relevant. 
In addition, assuming in each one of them an escape fraction of ionizing radiation higher than
zero (e.g., $>30\%$, \citealt{howard16}), 
they could contribute to induce fluorescence in the surrounding medium, generating diffuse Ly$\alpha$ nebulae.
Clearly, this  cannot be regarded as the only mechanism responsible for the Ly$\alpha$ nebulae detected so far.  
However, a significant contribution from proto-GCs or dwarf galaxies cannot be excluded, especially if their stellar 
emission is not detectable even in the deepest field surveys (such as, e.g., our GC1 source with magnitude 31.4). 
For instance, the origin of some of the Ly$\alpha$ nebulae discussed in 
\citet{vanz16c} is not well identified. 
Interestingly enough, a dozen proto-GC in action associated with a few high-z galaxies 
can produce $\simeq 20$\% of the
total Ly$\alpha$ luminosity observed in that case. Similarly, the clustering of faint objects around the main galaxy
might also increase the Ly$\alpha$ visibility during the reionization epoch \citep{castellano16a}.

Are some of the Ly$\alpha$ nebulae observed so far an indirect signature of the integrated 
contribution from elusive proto-GCs? Again, JWST can investigate this issue by providing deeper images 
than what available now, possibly revealing a multitude of currently undetected objects.

\section{Conclusions}

Deep observations provided by the Hubble Frontier Fields and deep
VLT/MUSE integral field spectroscopy, coupled with high-precision lens
models (based on tens of spectroscopic multiply imaged systems at
$3<z<6$), have allowed us to identify extremely faint objects within
the first two billion years after the Big-Bang, in a still unexplored region of
stellar mass and luminosity domains. These new regimes of
mass and luminosity are relevant for our understanding of the physics
of dwarf and globular cluster formation at high redshift, as well as
for the identification of sources possibly dominating the ionizing
background \cite[e.g.,][]{yue14, wise14}.

The main results are the following:

\begin{itemize}

\item{By taking advantage of (1) MUSE deep spectroscopy, (2) a
    detailed analysis of lensing magnification maps, (3) the spatial
    shapes of these selected sources (validated with lensing
    simulations), and (4) exploiting the Astrodeep HST photometry and
    SED fitting that includes nebular emission, we studied the
    faintest and most compact stellar systems at redshift $\sim$3 and
    $\sim$6 currently known ($-14 > M_{UV} > -17$), which are
    characterised by stellar masses in the range $10^{6}$\msun $ <$ M
    $< 20 \times 10^{6}$\msun\ and effective radii spanning the
    interval $\simeq16-150$ pc. }

\item{Two of our sources, GC1 and ID11, show stellar masses and
    star-formation rate densities consistent with the values expected
    in multi-population formation scenarios for GCs.  In particular,
    ID11 also shows a dynamical mass (derived from optical rest-frame
    spectroscopy) similar to the stellar mass, suggesting a negligible dark
    matter content in this system.
    In addition, the detection of high ionization lines like \civ\ and
    \heii\ (with a velocity dispersion of $\sigma_{v} \simeq 20$ \kms)
    suggests that hot stars are present, with an outflowing gas with
    velocity $\simeq 50$~\kms\ measured from line velocity offsets
    \citep{vanz16a}.  The same object also shows a low column density
    of neutral gas, lower than $10^{18.5}$cm$^{-2}$ (see
    \citealt{vanz16a}).  The other compact and dense object discovered
    at $z=6.145$, GC1, is perhaps the most intriguing source among
    those presented in this work, with an effective radius of $\simeq
    20$ pc and an intrinsic stellar mass of $2-4 \times 10^{6}
    M_{\odot}$.  Its properties are very close to those expected for a
    proto-GC.}

\item{We have also detected extremely faint knots in the system at
    redshift 6.145, whose de-lensed magnitudes are fainter than
    $32$. They are among the faintest objects at $z \sim 6$ ever found
    in any strongly lensed field.
The very nature of these extremely faint star-forming regions will be
better assessed with JWST and the extremely large telescopes.}
\end{itemize} 

The determination of sizes and physical properties (such as the
stellar mass) of large number of systems at $4 \le z \lesssim 8$ will
be greatly improved by means of JWST observations, that will perform
rest-frame optical spectroscopy and imaging with the NIRSpec and
NIRCam instruments, respectively.  In addition, the rest-frame
near-infrared wavelengths will be accessible with the JWST/MIRI
camera.\footnote{{\it https://jwst.stsci.edu/instrumentation}} Following the study of ID11 at $z=3.1169$ by means of VLT/X-Shooter near infrared
observations (see \citealt{vanz16a}),  JWST spectroscopy will allow
us to:

\noindent (1) investigate the nature of the
ionizing source from optical oxygen and Balmer lines ratios, in combination with ultraviolet features, 

\noindent (2) investigate the status of the interstellar
medium through line ratios in the optical and ultraviolet rest-frame 
(e.g., by calculating the O32 index, and looking for density and/or ionization-bounded signatures),

\noindent (3)  look for the presence of outflows possibly from the nebular emission of high
ionization lines (as we first attempted for ID11 with the \civ\ doublet),  

\noindent (4) perform direct estimates of the ionizing production rate
from the Balmer lines, as well as to investigate the escaping ionizing
radiation through indirect diagnostics calibrated at lower redshift
\citep{vanz16b}, specifically proposed for the characteristics of JWST
\cite[e.g.,][]{zackrisson13, zackrisson16}.  It is worth noting that
the current identification of photometric signatures of such nebular
lines imprinted in the Spitzer/IRAC bands is even more complicated by
the fact that at $5.5\lesssim z \lesssim 6.6$ both the 3.6$\mu m$ and
4.5$\mu m$ bands are polluted by Oxygen and \ha\ lines, respectively,
introducing a degeneracy that prevents any clear measure of their
equivalent widths \citep{smit15}.  JWST/NIRSpec
spectroscopy will remove this degeneracy.

Moreover, rest-frame optical and near infrared imaging with NIRCam and
MIRI will provide accurate estimates of the stellar masses and sizes
(now inferred from the ultraviolet light at $z\sim 6$).  The system at
$z \sim 6$ also represents an ideal target for integral filed
spectroscopy with JWST.  In particular, the $3'' \times 3''$ field of
view provided by NIRSpec-IFU will produce a cube of 900 spaxels which
contain GC1, D1 and \#22692 sources in a single shot, as well as the
possible stellar stream connecting all these features, enabling
kinematics studies in the optical rest-frame by using prominent lines
such as \oiidoub, \oiiidoub, \hb\ and \ha.

However, the study of ultraviolet absorption lines will require a good
detection of the continuum ($S/N>5$), achievable (thanks to
gravitational lensing) for objects brighter than $28$ at 1500\AA\
rest-frame and addressable with future extremely large telescopes
(e.g. E-ELT).

\section*{Acknowledgments}
The referee, Richard Ellis, is acknowledged for several constructive comments 
which significantly improved the presentation of the results. 
We thank Alvio Renzini, F. Ferraro and E. Dalessandro for several useful discussions.
C.G. acknowledges support by VILLUM FONDEN Young
Investigator Programme through grant no. 10123.  K.C. acknowledges
funding from the European Research Council through the award of the
Consolidator Grant ID 681627-BUILDUP.  Based on observations collected
at the European Southern Observatory for Astronomical research in the
Southern Hemisphere under ESO programmes P095.A-0653, P094.A-0115 (B)
and ID 094.A-0525(A). MM, AM and PR acknowledge the financial support
from PRIN-INAF 2014 1.05.01.94.02. MM acknowledges support from the
Italian Ministry of Foreign Affairs and International Cooperation,
Directorate General for Country Promotion.  We also thank Wouter
Karman for enlightening discussions over the last years.

\appendix
\section{Simulating ID11}
\label{simula}

 \begin{figure*}
\centering
\includegraphics[width=17cm]{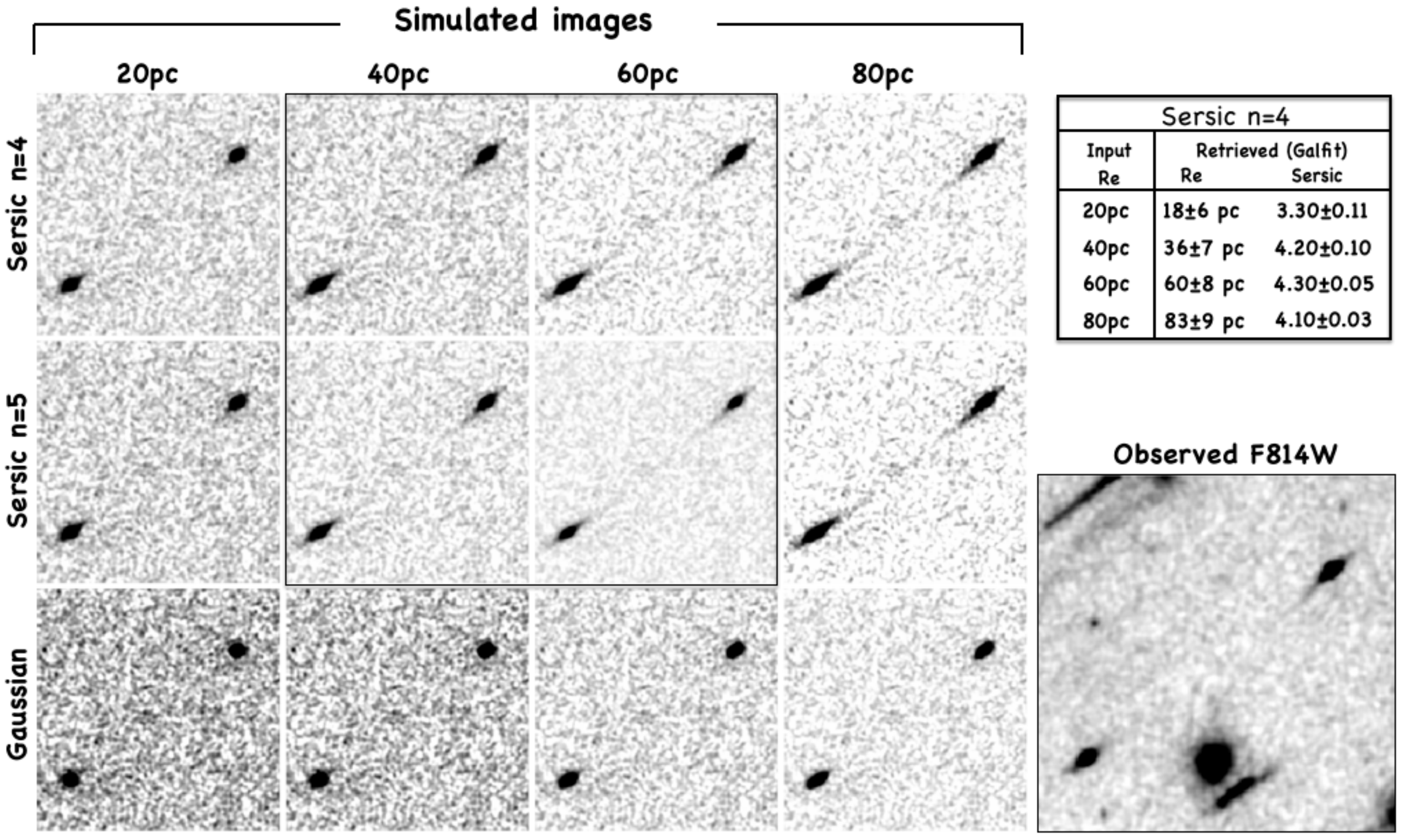}
\caption{Simulated  multiple images for ID11 are shown on the left side ($4\times3$ panels) by adopting different light profiles (S\'ersic, Gaussian) 
in the source plane 
and four effective radii (20, 40, 60, 80 pc). In the first two rows a S\'ersic index n=4 and n=5 are adopted, while in the bottom row
a two dimensional Gaussian is assumed. On the right side the F814W band image is shown. All the images have
been smoothed with a Gaussian filter with a radius of 3 pixel to highlight the tails of the elongated images. 
S\'ersic indexes $n\simeq 4-5$ and $R_{e} \simeq 40-50$ pc well reproduce 
the data (see text for details). The top-right table compares the input radii and S\'ersic index (only the $n=4$ case shown)
of simulated images in the source plane with those retrieved by performing {\tt Galfit} fitting on the lensed simulated images.}
\label{LensID11}
\end{figure*}

We have assessed the robustness and uncertainties of the structural
parameters reported above by performing end-to-end image simulations
with the software {\tt SkyLens} \citep{Meneghetti08,Meneghetti10}. As
outlined also by \citet{meneghetti16}, this code can be used to
simulate HST observations, including the lensing effects produced by
matter distributions along the line-of-sight to distant sources. In
the simulations presented here, we use our lens model of AS1063 to
lens some template sources placed at the redshift and predicted
position of ID11. In the following tests, we make the simplified assumption that the sources have circular shape.

We begin by performing simulations without including lensing effects. Since the light profile and the size of the 
sources is known, we can test if our method to measure these properties using the fitting software {\tt Galfit} is 
robust. We generate source templates assuming S\'ersic profiles with index in the range [0.5-10] and effective 
radius in the range [20-640] pc. \footnote{The following values have been used: 20, 40, 60, 80, 160, 320, 640 parsec.} 
We assume that the apparent magnitude of these sources in the F814W band 
is $m_{AB}=22.7$, which is chosen to provide an high S/N. 
We use {\tt Galfit} to retrieve the structural properties of the sources from the simulated observations, finding that 
profiles, radii, and magnitude are correctly recovered only in the case of sources with $R_e=640, 320$ and 160 pc,
the latter radius corresponds to 0.7 pixels at the redshift of ID11.
When a smaller $R_e$ is used, the profile parameters are only marginally recovered, showing the limits of the PSF 
de-convolution implemented by {\tt Galfit}. This latter cases ($R_e < 100$pc) corresponds to $R_e<0.5$ pixels.
\citet{peng10} shows that objects with $R_e$ smaller than half a pixel are not spatially resolved and 
severe systematics errors dominate any fit\footnote{Clearly this depends also
on the S/N of the image to fit and the quality of the PSF used. We limit our analysis to our PSF and for a bright
object with magnitude 22.7, implying that for fainter not resolved objects the systematic errors are even more severe.}.    

Then, we repeat the above experiment by simulating the lensing effects produced by AS1063. 
Figure~\ref{LensID11} compares the results of the simulations with the real images of ID11 observed in the
F814W band. While a Gaussian profile (S\'ersic index $n=0.5$) is clearly inconsistent with the morphology of the 
images A and B of ID11 for any value of $R_e$,  source models assuming S\'ersic profiles with larger $n$ reproduce 
remarkably well the observed distortions, both in terms of their amplitude (i.e. the magnification) and direction. 
In particular the best agreement is found for sources with $n\sim4-5$ and $R_e\sim 40-60$ pc 
(see again Figure~\ref{LensID11}).

{\tt Galfit} fitting is performed on all the simulated images and the recovered parameters are compared to those
of the input sources. In this specific case, the presence of two multiple images with very similar magnification (A and B)
allow us to further check the variation in the {\tt Galfit} measurements. In the cases $n=4$ and effective radii [20 - 80] pc 
we recover all the input parameters reasonably well. For example, in the case of the smallest source, $R_e=20$ pc, 
the measured effective radius is $R_e=1.28\pm0.25$ pixels (in the tangential direction), that corresponds to 
$R_e=1.28 \times 0.03 \times 7650 / \mu_{T}$ = 18.1 pc. 
Similarly, the measured structural properties are consistent with the
input source models for the other values of $n$ and $R_e$ (see the summary table in Figure~\ref{LensID11}). We can
therefore conclude that: (1) simulations validate the adopted method
to infer the half light radii and (2) objects with effective radii as
small as 20 pc are recoverable, assuming they are tangentially
magnified consistently to images A and B of ID11.

\end{document}